\documentclass[fleqn,usenatbib]{mnras}
\usepackage{float}
\usepackage{amsmath}
\usepackage{amssymb}
\usepackage{rotating}
\usepackage{tabularx}
\usepackage{lscape}
\usepackage{xr}

\usepackage[font=footnotesize,labelfont=bf]{caption}

\begin{document}
\include{journaldefs}
\title[\bf  VPHAS+ in the third quadrant]{\bf 
Uniformly-calibrated VPHAS+ photometry in the third quadrant of the Galactic plane
}
\author[J. E. Drew, et al]{
{\parbox{\textwidth}{J. E. Drew$^{1}$\thanks{E-mail: j.drew@ucl.ac.uk}, R. Greimel$^2$, J. Eisl\"offel$^3$, R. Raddi$^4$, N. J. Wright$^5$}
}\\ \\
$^{1}$Department of Physics \& Astronomy, University College London, Gower Street, London, WC1E 6BT, UK \\
$^{2}$RG Science, Schanzelgasse 17, A-8010 Graz, Austria \\
$^3$Thüringer Landessternwarte, Sternwarte 5, 07778 Tautenburg, Germany\\
$^4$Universitat Politècnica de Catalunya, Departament de Física, c/ Esteve Terrades 5, E-08860 Castelldefels, Spain\\
$^5$Astrophysics Research Centre, Keele University, Keele ST5 5BG, UK\\
}

\maketitle

\begin{abstract}
The southern Galactic plane has been mapped at optical wavelengths and at under one-arcsecond angular resolution by the VST Photometric H$\alpha$ Survey of the Galactic plane and bulge (VPHAS+). Anticipating the release of a uniform photometric calibration of the entire survey, we examine the properties of VPHAS+ $ugriH\alpha$ photometry of $r < 19$ mag. point sources  in the third Galactic quadrant (longitudes $210^{\circ} \lesssim \ell < 260^{\circ}$).  We compare our interim calibration in $gri$ with that of Pan-STARRS, the DECam Plane Survey (DECaPS-2) and Skymapper. 
We use the comparisons to identify small $gri$ photometric offsets.  Corrections to the $H\alpha$ and $u$ magnitude scales are determined via comparison with synthetic photometry. VPHAS+ and its northern counterpart, the INT Galactic Plane Survey (IGAPS), are shown to closely align, where they overlap across the celestial equator.  Aided by Gaia Data Release 3, SIMBAD, and specialist catalogues, we present selections of: A stars; sub-luminous stars; intrinsically-red luminous stars; young stellar objects; emission-line and OB stars. Attention is drawn to stellar variability as a contaminant in selecting emission line objects via $(r - H\alpha)$ excess.  It is argued the $(r - i, r - H\alpha)$ plane is the better choice for this selection than $(g - i, r - H\alpha)$.   Using A stars to map extinction, we trace the main run of dust obscuration, situated at mainly negative Galactic latitudes.
Like the dust, OB and emission line stars are more frequent below the Galactic equator: at heliocentric distances of up to $\sim$7~kpc, these stars' distribution fit in with the known warping of the Galactic plane.  An overdensity of B stars, several degrees across and potentially in the Outer Arm, is found around $(\ell,b) = (212^{\circ}.0,-0.6^{\circ})$.
\end{abstract}



\section{Introduction}

The southern hemisphere offers a view of around two-thirds of the Milky Way's stars. Most of the directly observable baryonic mass of our own galaxy is concentrated in the disk occupying its equatorial plane, and in towards the Galactic centre, in the southern constellation of Sagittarius.   The mass of the Galaxy is in three forms: stars, gas and dust.  The relative abundance of dust in the Galactic plane, peaking in the molecular ring at a Galactocentric radius of $\sim 3kpc$, introduces a large irregular distribution of extinction across the plane that reduces the optical visibility of its stellar content. In short, along with more stars, the southern plane hosts the largest dust columns. It is a place of contrasts, offering long sparsely-populated sight lines outside the Solar Circle in the third quadrant, as well as views of the denser inner disk and of the dynamically-distinct centre and bulge.  The Galactic plane continues to be important as the nearest, best-resolved environment in which the processes of star formation, stellar and galactic evolution play out. 

A practical consequence of the plane's dust content has been that surveys of the Galactic plane and bulge have turned to near-infrared (NIR) and longer wavelengths -- particularly where the main science driver has been studies of the earlier stages of star formation \citep{Glimpse2003,VVV2012, Roman2023}. However, the optical domain still has much to offer.
In terms of impact on the evolving Galactic environment, the short lifetimes and greater luminosities of higher-mass stars make them important as agents for change, mediated by their radiation and steady or explosive mass loss.  For these stars, the optical/ultraviolet is where the Planck maxima in their spectral energy distributions (SEDs) are located, with the result that photometric - and spectroscopic - discrimination of effective temperature is easier in the optical than at longer wavelengths.  Only the SEDs of M stars and cooler brown dwarfs peak in the NIR domain. Compact late stages of stellar evolution and their ejecta also continue to benefit from optical observation.  
There is a great legacy of mature CCD detector technology that continues to enable ever more ambitious optical projects: the recently-complete Gaia astrometric mission \citep{Gaia2016} and the forthcoming Rubin Observatory Legacy Survey of Space \& Time \citep{LSST2019} are prominent examples of this.

Atomic physics has placed the outstanding diagnostic of diffuse ionized gas, the H$\alpha$ transition, in the red-optical domain.  Exploitation of this feature is a central motivation of the VST Photometric H$\alpha$ Survey of the Southern Galactic Plane and Bulge \citep[VPHAS+,][]{Drew2014} -- the subject of this paper.  Coming after counterpart surveys in the northern hemisphere, namely IPHAS\footnote{The INT photometric H$\alpha$ survey of the northern Galactic plane} \citep{Drew2005} and UVEX\footnote{The UV-excess survey of the northern Galactic plane} \citep{Groot2009}, VPHAS+ was set up to combine narrow-band H$\alpha$ and Sloan broad-band imaging that would provide reliable colour-colour data, capable of serving a broad range of Galactic-plane stellar and ISM astrophysics. The inclusion of H$\alpha$ narrow band and Sloan $u$ band are of particular diagnostic importance for stellar analysis, with the former providing crude, but extremely resource-efficient, spectroscopy and the latter, key to picking out UV-excess objects.  

The aim of this paper is to provide a VPHAS$+$ primer, ahead of the full release of uniformly calibrated VPHAS+ point-source photometry, by examining the characteristics and utility of the photometry with reference to a relatively uncomplicated section of the footprint.  We choose for this purpose, the Galactic longitude range from $208^{\circ}$, the beginning of the VPHAS$+$ footprint near the celestial equator, up to  $\ell = 260^{\circ}$ -- all contained within the third Galactic quadrant, and lying outside the Solar Circle. Here, there is not much of an extinction accumulation, near the Sun, within the $-5^{\circ} < b < +5^{\circ}$ survey band \citep[see][and other extinction maps cited therein]{Edenhofer2024}. In principle this eases making and understanding comparisons with other published overlapping photometric surveys offering data in, nominally, the same filters. 

The most nearly comparable southern plane survey data, already available, are from Pan-STARRS 1 \citep[][hereafter PS1]{Chambers2016} and DECaPS-2 \citep{Saydjari2023}.  The median angular resolutions achieved in Sloan $g$, $r$ and $i$, the three bands in common with VPHAS+, span the range 1.1--1.4 arcsec.  The VPHAS$+$ data are usually better resolved than this: for the same filters, the medians range from 0.7 to 0.85 arcsec.  Both PS1 and DECaPS-2 reach down to at least $\sim$22nd magnitude in these same bands, while the faint limit of VPHAS$+$ is about a magnitude brighter. We also compare with Skymapper DR4 \citep{Onken2024}.  Although the angular resolution Skymapper offers is coarser by a factor of a few, and the data are not so deep, the survey usefully overlaps all the VPHAS+ footprint.      
Our last comparison is with IGAPS \citep{Monguio20} in the overlap region across the celestial equator.

In the second half of this paper we explore ways to build specialist selections of objects for science exploitation.  The starting points for this are cross-matches with Gaia DR3, the SIMBAD database and a range of specialist catalogues..  In doing this, we draw on the experience with selections of the same groups before now using smaller VPHAS+ sky areas \citep{Kalari2015,Raddi2016, MMS2017, Drew21} and also the IGAPS constituent surveys \citep[e.g.][]{Drew2008, Wright2008, Verbeek2012, Raddi2013}.  In this process, more of the properties of the Galactic disc in the third quadrant are exposed, and magnitude corrections arising from the inter-survey comparisons can be checked.

The contents of this paper are as follows.  Section~\ref{sec:data} provides an outline of the VPHAS+ survey, its data and data products.  Comparisons with other overlapping photometric surveys, including IGAPS, and with synthetic photometry, are presented in Section~\ref{sec:comparisons}.  These are all conducted with magnitudes expressed in the AB system.  Through these comparisons we refine the zeropoints in the VPHAS+ bands, and examine how uniform the survey is across the third quadrant.  

The rest of the paper illustrates a range of applications, relating to specific classes of star: of necessity, this needs to be carried out with the survey magnitudes re-expressed in the Vega system. Conversions to Vega are identified in Section~\ref{sec:ABtoVega}.  
Section~\ref{sec:astars} begins the demonstration science, with an A-star selection for use as a mapping tool. Section~\ref{sec:SIMBAD} covers subluminous stars, intrinsically red stars and young stellar objects. This is followed by a selection of emission line candidate objects in Section~\ref{sec:emstars}.  Here, the problem posed by stellar variability is brought into the foreground.  The last object class to be considered are the massive OB stars, in Section~\ref{sec:OBstars}. The paper ends, in Section~\ref{sec:discussion}, with a concluding discussion. 

\section{VPHAS+ Data}
\label{sec:data}

\subsection{Observations and data processing}
\label{sec:observations}

The VST observations contained within the VPHAS+ database were collected from the end of 2011 up to August 2018.  The survey's design and reduction pipeline were described by \cite{Drew2014}.  The survey observing blocks (OBs) were of two main types that split the five photometric filters between them: the blue OBs were set up to capture the $u$, $g$ and $r$ bands for three (usually) contiguous fields; the red OBs included $r$ again, along with narrow-band $H\alpha$\footnote{This filter is named NB 659 by ESO} and $i$, and again three fields were covered per OB.  For any given 1 sq.deg. field within the survey footprint, the dates of acquisition of the blue and red data are almost never the same, and may even be years apart.  This strategy was adopted, at the request of the European Southern Observatory (ESO), to improve access to and the efficiency of the telescope, exploiting the fact that the sky conditions needed for good-quality red filter data are less demanding.  It happened, also, to replicate the distinct schedules of the IPHAS \citep{Drew2005} and UVEX \citep{Groot2009} surveys of the northern Galactic plane that were merged in IGAPS \citep{Monguio20}.  

The presence of $r$ filter observations in both the blue and red OBs allows longer-term secular variability of point sources to be clearly identified (a well-known feature of long period variables, for instance).  OB duration was required by the observatory to be less than an hour, with each OB structured to minimise filter changes.  This meant that the three fields observed were cycled through each of the three filters, creating an elapsed time between the first and last filter exposure observed per field of 30 to 45 minutes.  Hence any time variability on timescales of $\sim$ an hour or less is effectively scrambled.  Whilst variability is more the rule than the exception among stars generally, such rapid variation at the measurable $>0.01$ mag. scale is a minority phenomenon \citep{Heinze2018}.

The originally planned VPHAS+ footprint included the Galactic latitude range $-5^{\circ} < b < +5^{\circ}$ at all Galactic latitudes in the southern hemisphere. However, due in large part to overrunning the duration of research support, data collection ended in August 2018, after 6.7 years of survey operation, with parts of the 4th quadrant unobserved.  The focus in this work is on the third quadrant where coverage is complete.  

Full details of the observing pattern in terms of offsets per filter and exposure times are available in the original survey paper \citep{Drew2014}.  Here, it is important to note that the merged catalogue in use gives a single set of magnitudes per source, after averaging the aperture photometry from all available detections per source per filter. This is being followed up by the construction of a uniform calibration to be presented in full and final form, by Greimel et al (in preparation).  The penultimate version in use here already accounts well for the photometry outside the Solar Circle, thanks to the low-to-moderate dust columns encountered there.

The key feature of the uniform calibration (final and the version used here) is that it uses space-based Gaia EDR3 $G$, $BP$ and $RP$ photometry as the presumed uniform reference to which VPHAS+ magnitudes should be aligned across the footprint.  These comparisons are conducted at the OmegaCam CCD level ($7.5 \times 15.0$ arcmin$^2$ on sky).  Merging then takes place, in which offset frames and other overlaps are brought to a common photometric level per filter: catalogues are generated on a Galactic Coordinate grid such that there is finally a band-merged catalogue per square degree.  These are combined here to span a sky area of $\sim$500 sq.deg.
The AB system of magnitudes is in place, through inheritance from the pipeline zeropoints determined using APASS DR9\footnote{See https://ui.adsabs.harvard.edu/abs/2015AAS...22533616H}. 

\subsection{The merged catalogue}

\begin{figure*}
\begin{center}
\includegraphics[width=2.1
\columnwidth]{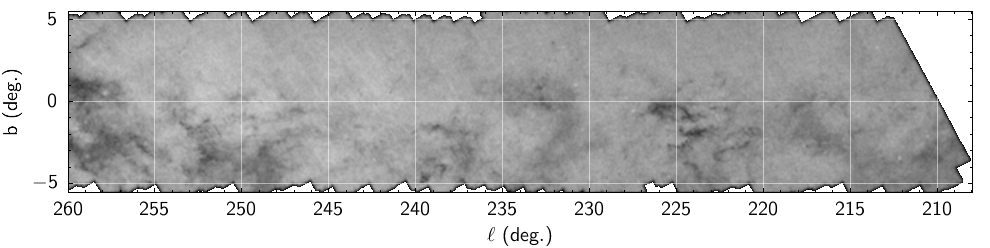}
\caption{
Section of the VPHAS+ footprint examined in this study, shown as a stellar density map.  The grayscale applied is shaded according to the square root of the density.  The darker areas pick out the parts of the region with lower stellar density, where there is a greater accumulation of extinction.  The level of contrast present is of the order of a factor of 10.  In the fourth and first quadrants of the Galactic plane, density variations are much more pronounced. 
}
\label{fig:density_map}
\end{center}
\end{figure*}

The purposes of this paper are best served by a catalogue, Q3-bright, spanning the Galactic longitude range from $\sim210^{\circ}$ to $260^{\circ}$, that is limited to the better quality data associated with brighter magnitudes and star-like sources with no or little impact from blending on the photometry\footnote{'Star-like' and 'unblended' is assured by constraints on the catalogue columns 'mergedclass' ($<2.5$), and 'errbits' ($< 3$ in $g$, $r$ or $i$).  Errbits = 0, indicates an unblended  source, while errbits = 2 indicates pipeline deblending in which the source tagged is the brighter component.}.  The magnitude cuts applied are that the $r$ magnitude lies in the range $13 < r < 19$ mag., where $r$ is a mean magnitude, $(r_r + r_b)/2$ in which $r_r$ and $r_b$ distinguish the red-OB and blue-OB $r$ magnitude measurements, respectively. This leaves out the much larger number of detections fainter than the $r = 19$ limit \citep[see][]{Drew2014}.  The number of sources in Q3-bright is 11.42 million.  Their sky distribution is shown as a density map in Fig.~\ref{fig:density_map}.  With increasing longitude there is a slow rise in stellar density (towards lighter grey at the left end of the plot), whilst there is evidently more obscuration affecting the negative latitudes - an effect that becomes even more marked near the $\ell = 260^{\circ}$ boundary and spreads to positive latitude.  Indeed, the choice of $\ell = 260^{\circ}$ as the upper boundary was prompted by the growing impact of extinction as the fourth quadrant, and the Carina Arm tangent, are approached.

\begin{figure}
\begin{center}
\includegraphics[width=0.9\columnwidth]{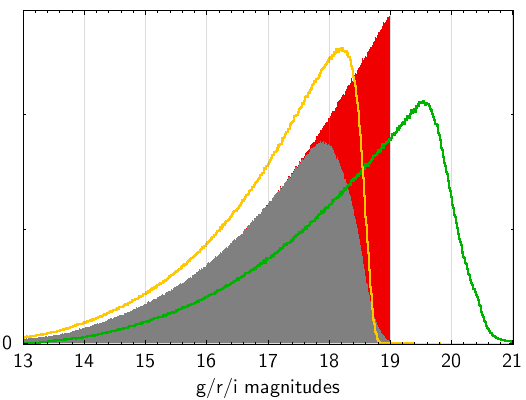}
\caption{Magnitude distributions in the $g$, $r$ and $i$ bands.  The primary selection of point sources for inclusion in this study requires $r = 0.5(r_r + r_b) < 19.0$ mag., good point-source morphology, and little/no photometric impact from  blending with neighbours in $g$, $r$ or $i$ (Q3-bright). The resultant $r$ AB-magnitude distribution is in solid red.  The $g$ and $i$ magnitude distributions are outlined by the green and yellow lines, respectively.  The solid grey distribution (on top of the red) is the $r$ distribution after imposing the further constraints that $g$, $r$ and $i$ pipeline errors are less than 0.01.  The vertical scale is linear source counts.
}
\label{fig:mags}
\end{center}
\end{figure} 

\begin{figure}
\begin{center} 
\includegraphics[width=0.9\columnwidth]{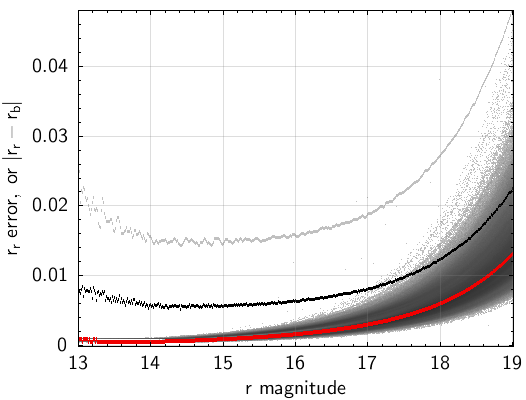}
\caption{Uncertainty in $r$ magnitude over the range $13 < r < 19$~mag.  It is presented in two ways: as pipeline-reported random error (distribution shown in a density grayscale, with the median in red), and as $|r_r - r_b|$ -- a more directly empirical measure that brings into account systematic effects as well.  The latter is represented by the 68th and 95th percentiles of the distribution, respectively in black and lighter grey.
}
\label{fig:uncertainty}
\end{center}
\end{figure} 

Whilst the magnitude distribution in $r$ has been cut sharply at 19th, the distributions in $g$ and $i$ are left to roll off smoothly at the faint end (Fig.~\ref{fig:mags}).  As the common property of stars in the Galactic plane is that they are either reddened by extinction or intrinsically red (or both), it follows that the peak of the $i$ distribution is brighter than 19th mag., and that the peak of the $g$ distribution is fainter.  Revisualising these magnitude distributions as colours, we find peak $(r_r - i, g - r_b)$ occurs at $(0.3,0.8)$ - the colours of an unreddened early K star in the AB system \citep[Table 1 in][]{Fukugita2011}.  

The magnitudes provided 
are determined by aperture photometry and are accompanied by pipeline-determined errors.  These are essentially Poisson errors that represent only part of the total budget, particularly at brighter magnitudes: this is illustrated by Fig.~\ref{fig:uncertainty} in which percentiles of the measured differences $|r_r - r_b|$ are compared with the full distribution of pipeline errors in $r_r$ (shown as a density plot).  The actual differences between the red and blue $r$ magnitudes, from exposures obtained at different epochs, show a bottoming out of the 1-sigma scatter at around 6 milli-magnitudes in the $14 < r < 16$ range -- even as the formal pipeline errors drop to much lower levels.  At the same time, the $r_r$ and $r_b$ scales are aligned by design: the mean $(r_b - r_r)$ is 0.00018 down to 19th magnitude, while for $r < 18$ it is only $\sim 0.00005$.

\begin{figure}
\begin{center} 
\includegraphics[width=0.9\columnwidth]{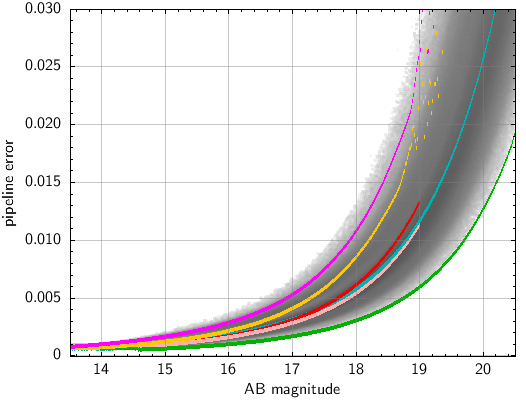}
\caption{Median pipeline errors for all bands as a function of magnitude.  The curves are colour-coded as follows: NB659/H$\alpha$ -- magenta; $i$ -- deep yellow; red $r$ -- red; blue $r$ -- pink, $g$ -- green; $u$ -- cyan.  For the $u$ band the full distribution of errors is shown as a grayscale density plot.  
}
\label{fig:errors_by_filter}
\end{center}
\end{figure}

Fig.~\ref{fig:errors_by_filter} shows how the pipeline errors compare in the different survey filters.  The NB659/H$\alpha$ narrow-band errors are highest, because the narrowness of the band is not fully compensated for by increased exposure time.  The $i$ band errors are next highest, while the $g$ filter is associated with the lowest -- as this is where the detector response, peaks.  The red and blue $r$ errors are much the same, with blue $r$ running a little lower, due to the typically darker sky required for blue observing blocks.  Since most detected objects in the survey are red or reddened, the general pattern per object is that the errors in $g/r/i$ are comparable, while the $u$ filter errors are typically much larger, since $u$ magnitudes are commonly fainter by 1--2 mag. (for $u$ and H$\alpha$, it is impractical to expose long enough to bring the errors down to match those in $gri$).   

\section{Comparisons with other surveys and synthetic photometry} 
\label{sec:comparisons}

In this section we make a series of photometric comparisons, with the main aim of identifying zero-point adjustments in the VPHAS+ survey bands. To this end, a further set of quality cuts is applied in order to sharpen up Q3-bright.  The most critical comparisons are in $g$, $r$ and $i$. Hence, the cuts imposed are that the pipeline-provided errors on $g$, $r$ and $i$ magnitudes should be less than 0.01 mag., and that, in these same bands, there is no source blending flagged.  These extra constraints reduce the number of sources to 6.62 million.  It also moves the peak of the $r$ magnitude distribution from 19.0, down to 17.9 mag. (see Fig.~\ref{fig:mags}) -- more and more stars drop out as magnitude increases beyond this. There is a commensurate shift to brighter magnitudes in the $g$- and $i$-band distribution peaks.

\subsection{The common colour-colour diagram: (g-r) versus (r-i)}
\label{sec:phot-colours}

\begin{figure*}
\begin{center} 
\includegraphics[width=0.86\columnwidth]{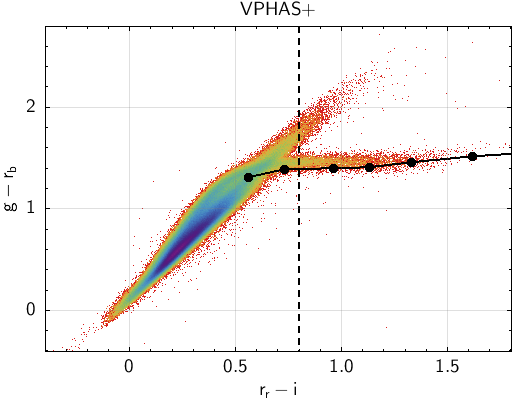}
\includegraphics[width=0.86\columnwidth]{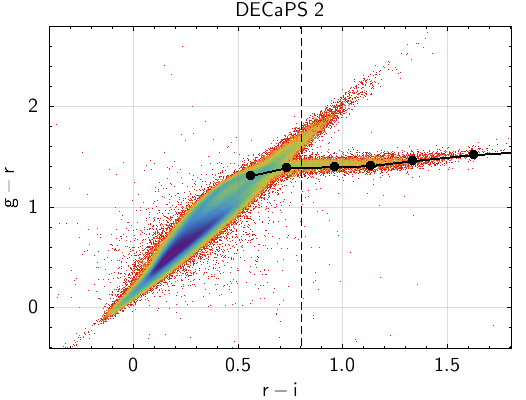}
\includegraphics[width=0.86\columnwidth]{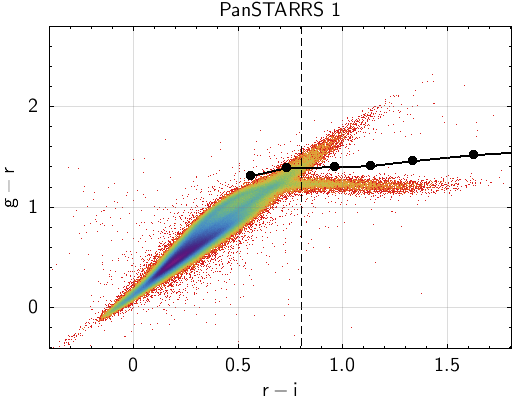}
\includegraphics[width=0.86
\columnwidth]{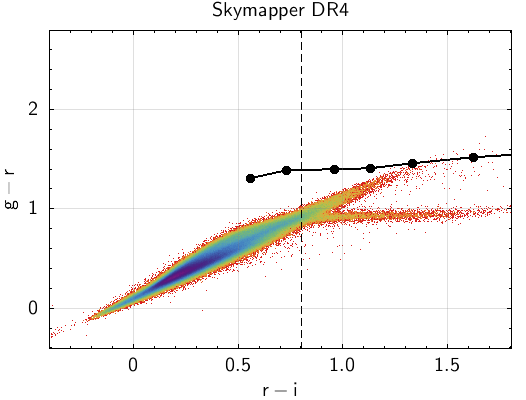}
\caption{
Colour-colour $gri$ diagrams obtained from the VPHAS+, DECaPS-2, Pan-STARRS 1 and Skymapper DR4 survey catalogues, for the sky area $240^{\circ} < \ell < 245^{\circ}$, $-5^{\circ} < b < +5^{\circ}$. The distributions are coloured according to the logarithm of point density (low density, red; high density, violet).  The linked black data points mark the AB colours of M dwarfs as determined from the Sloan Digital Sky Survey by West et al. (2011). The vertical dashed line at $r - i = 0.8$ is for reference against the $r - i$ axis. The VPHAS+, DECaPS-2 and PS1 datasets have been cleaned by requiring magnitude errors $< 0.01$ (leaving of order 1 million stars in each survey excerpt).  For Skymapper, this error limit was relaxed to $< 0.02$ (641k stars).  
}
\label{fig:gri_diagrams}
\end{center}
\end{figure*} 

\begin{figure*}
\begin{center}
\includegraphics[width=1.5\columnwidth]{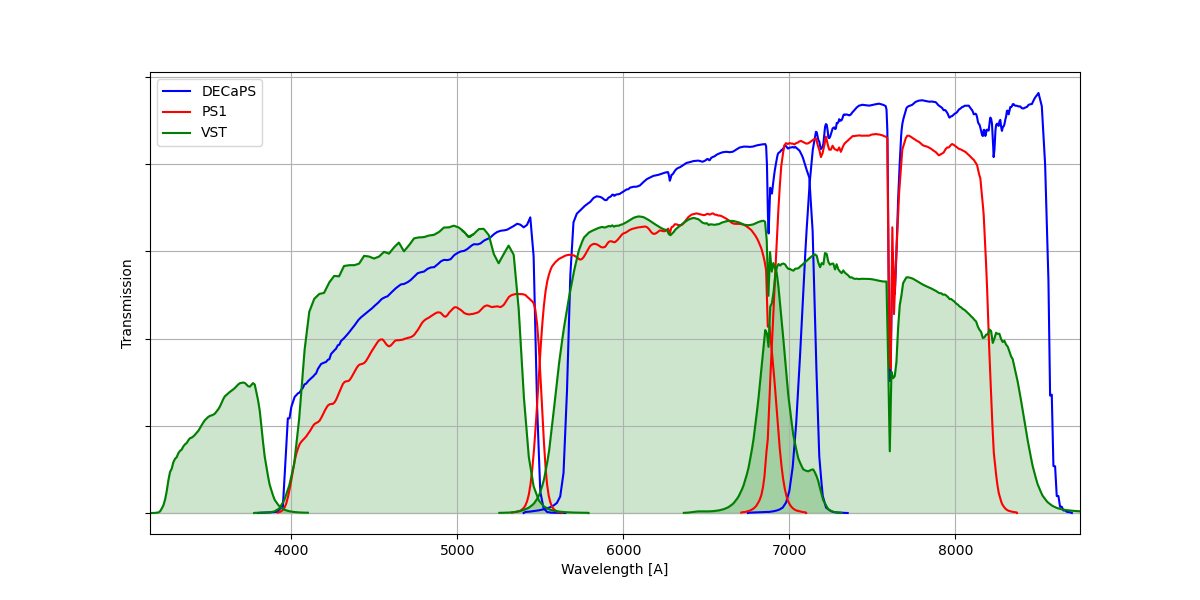}
\caption{
Filter profiles for the VPHAS+ (green, shaded), PS1 (red) and DECaPS-2 (blue) surveys.  All profiles are shown with atmospheric absorption and CCD detector response included. The scaling of the filter sets with respect to each other is arbitrary.  Within each filter set, the relative scalings are meaningful, serving to show that the PS1 and DECaPS-2 cameras are red sensitive, while OmegaCam -- used by VPHAS+ -- is more blue sensitive.  The $gri$ profiles are shown for all three surveys, while $u$ is also included for VPHAS+.
}
\label{fig:transmission}
\end{center}  
\end{figure*}

Our aim is to see how the VPHAS+ broadband magnitude scales measure up against existing alternatives. A factor in this will be the known and unavoidable problem of the differing transmissions of filters used in different surveys going by the same band name: quite modest differences in transmission profile, combined with the wide spread among intrinsic and/or reddened stellar colours, can result in noticeably different photometric behaviour. 

First, we compare VPHAS+ with the public survey data of PS1 and DECaPS-2, and Skymapper DR4 via the $(g-r, r-i)$ colour-colour plane (shortened to $gri$ plane, hereafter)\footnote{Similar contractions will be used for other colour-colour planes: $(r-H\alpha,r-i)$ becomes $riH\alpha$, $(u-g,g-r)$ becomes $ugr$ and $(g-i,r-H\alpha)$ becomes $girH\alpha$.}, as constructed for the area of sky, common to all four surveys  ($240^{\circ} < \ell < 245^{\circ}$, $|b| < 5^{\circ}$). 
Fig.~\ref{fig:gri_diagrams} presents the three colour-colour diagrams.  The most obvious point of contrast between them is brought out by the position of the horizontal M-dwarf spur off the main stellar locus.  To emphasise this, the AB colours of M dwarfs reported by \cite{West2011} -- as measured from Sloan Digital Sky Survey (SDSS) data -- are over-plotted in each panel as a common reference.  They largely coincide with the spurs in the VPHAS+ and DECaPS-2 diagrams, while in PS1, and even more so in Skymapper DR4, they sit above.  This pattern is the consequence of the differences in $g - r$ wavelength baseline.  

To reveal what is going on, the transmission profiles for the VPHAS+, DECaPS-2 and PS1 $gri$ filter sets are compared in Fig.~\ref{fig:transmission}. The key difference is that the PS1 $r$ (and $i$) filter transmissions start and end at bluer wavelengths than do their DECaPS-2 and VPHAS+ counterparts, while the $g$ filters are more nearly aligned.  Filter profiles for Skymapper have been presented by \cite{Onken2024} in their Fig. 1.  The $(g-r)$ range of  the Skymapper data shown in Fig.~\ref{fig:gri_diagrams} is the most compressed, in response to the significant overlap of the extended red roll-off of the survey's $g$-band transmission, with a slightly bluer $r$ band.   The $(r-i)$ dynamic ranges are more nearly the same in all four surveys -- it takes closer inspection of Fig.~\ref{fig:gri_diagrams} to notice the differences.  The greatest stretch in $(r-i)$ is seen in the Skymapper diagram, again thanks to its bluer $r$ filter, set alongside an $i$ band centering similarly to the DECaPS-2 $i$ band.

\subsection{Photometric scale comparisons with PS1, DECaPS-2 and Skymapper DR4}
\label{sec:phot-scale}

\begin{figure}
\begin{center} 
\includegraphics[width=0.9\columnwidth]{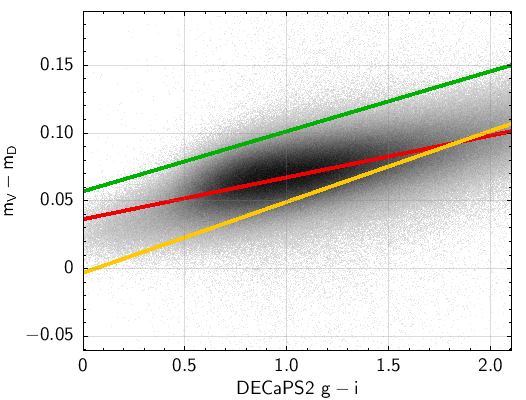}
\caption{Differences between VPHAS$+$ and DECaPS-2 $gri$ magnitudes ($m_V - m_D$) as a function of DECaPS-2 $(g-i)$ colour.   These have been computed for 2.3M cross-matched stars, in the region ($\ell > 240^{\circ}$, $-5^{\circ} < b < +5^{\circ}$). Linear fits are shown for each of the filters: green for $g$, red for $r$ and yellow for $i$.  The full $r$ distribution, shaded for density (square-root grey scale) is shown. 
Only stars with errors less than 0.01 mag. in all 3 bands in both surveys are included.  
}
\label{fig:VPHASvDECaPS}
\end{center}
\end{figure}

In principle, the absolute level of the VPHAS+ photometric scale can be compared with any of PS1, DECaPS-2 and Skymapper, where they overlap. The PS1 overlap runs between $\ell \sim 208^{\circ}$ and $\ell = 245^{\circ}$.  For DECaPS-2, the overlap begins at $\ell = 240^{\circ}$ and runs beyond the $\ell = 260^{\circ}$ limit set for this paper.  The Skymapper overlap is total, spanning the full longitude range.  Ideally, cross-matching with these surveys would reveal the same set of $gri$ global zeropoint shifts, to within the errors.  For this to happen all three surveys would need to be photometrically-aligned.  DECaPS-2 has been aligned with PS1 off the Galactic plane \citep[see Fig.13 in][]{Saydjari2023}, but it may not hold in the plane, where extinction and stellar population changes can combine with filter differences between the two surveys to introduce subtle shifts.  

There exist small corrections to PS1 $gri$ zeropoints that arose from a careful re-evaluation of CalSpec standards by \cite{Scolnic2015}.  These amount to positive magnitude increments of 0.020, 0.033 and 0.024 for $g$, $r$ and $i$ respectively -- all with uncertainties of 0.008 mag.  \cite{Magnier2020} describe their implementation within the second data reduction (DR2) of PS1.  Cross-matching tests indicate that these corrections are included in the DR1 catalogue, that was made available to CDS Strasbourg in 2017\footnote{Table II/349, which can be queried via https://vizier.cds.unistra.fr/viz-bin/VizieR-2}. This is our source for PS1 data. The corrections are not included by default in the DECaPS-2 catalogue, but are mentioned as appropriate in the introduction to the catalogue web pages\footnote{http://decaps.skymaps.info/catalogs.html}, for maintaining photometric alignment with PS1  \citep[see Section 6.2 in][]{Saydjari2023}.   
Accordingly we always apply the \cite{Scolnic2015} corrections to DECaPS-2 magnitudes before using them.  In contrast, the photometric calibration of Skymapper DR4 described by \cite{Onken2024} is not linked to PS1: to achieve uniformity, they make use of synthetic photometry computed from Gaia DR3 prism spectra.

\begin{table*}
\caption{Coefficients, $c_0$ and $c_1$, of the linear colour equations obtained for the comparisons between Q3-bright and each of DECaPS-2, PS1 and Skymapper DR4. The RMS variation of the magnitude differences, $\sigma$ is also provided.}
{\centering
\begin{tabular}{lccccccccc}
\hline       
     & \multicolumn{3}{c}{$n_V - m_D$} & \multicolumn{3}{c}{$m_V - m_P$} & \multicolumn{3}{c}{$m_V - m_S$} \\
 $\ell$ range: & \multicolumn{3}{c}{$240^{\circ} - 260^{\circ}$} & \multicolumn{3}{c}{$210^{\circ} - 245^{\circ}$} & \multicolumn{3}{c}{$210^{\circ} - 260^{\circ}$} \\    
Band  & $c_o$ & $c_1$ & $\sigma$ & $c_o$ & $c_1$ & $\sigma$ & $c_o$ & $c_1$ & $\sigma$ \\ 
\hline     
g & 0.0569 & 0.0442 & 0.019 \ \ & 0.0238 & 0.0856 & 0.016 \ \ & 0.0491 & 0.2278 & 0.024 \\
r & 0.0367 & 0.0307 & 0.017 \ \ & 0.0064 & $-$0.0297 & 0.014 \ \ & 0.0134 & $-$0.0630 & 0.015 \\
i & $-$0.0034 & 0.0526 & 0.019 \ \ & $-$0.0160 & $-$0.0018 & 0.014 \ \ & $-$0.0179 & 0.0476 & 0.015 \\ 
\hline
\end{tabular}
 }
\label{tab:colour_eqns} 
\end{table*}

We can now investigate $gri$ magnitude differences as functions of colour between Q3-bright and the three overlapping surveys.  We choose $(g - i)$ as the reference colour, emulating the practice of \cite{Schlafly2018} and \cite{Saydjari2023} in calibrating DECaPS: the fully-encompassing wavelength baseline is helpful.  We fit linear trends of the form
\begin{equation}
   m_V  - m_X = c_0 + c_1.(g - i)_X    
\end{equation}
in which $X$ identifies the comparison survey ($X = D$ for DECaPS-2, $P$ for PS1 and $S$ for Skymapper DR4).  The fits
are computed over the colour range, $0 < g - i < 2.5$, that always captures the distribution peak and excludes extremely blue and red stars.
The linear coefficients for each of the $g$, $r$ and $i$ bands, and the associated fit RMS values are given in Table~\ref{tab:colour_eqns}.

The best-fitting linear dependences on $(g - i)$ colour obtained from the comparison between Q3-bright and DECaPS\_2 are plotted in Fig.~\ref{fig:VPHASvDECaPS}.  The similarity between the VPHAS+ and DECaPS-2 $gri$ planes shown in Fig.~\ref{fig:gri_diagrams} connects with the way in which all $m_V - m_D$ rise with increasing $g-i$, without much divergence.  The pattern is not as tidy in the comparisons with PS1 and Skymapper DR4, where mixes of positive and negative gradients ($c_1$) apply.

\begin{table*}
\caption{Median $gri$ magnitude offsets computed using photometrically high-quality stars in common between Q3-bright, PS1 and DECaPS-2 and Skymapper DR4.  To align the Q3-bright photometric scale with these surveys separately, these quantities would be added to the Q3-bright zeropoints or, equivalently, subtracted from magnitudes.  The final column gives the corrections we apply to the Q3-bright catalogue, formed from the overlap-area weighted means (the weights used are given in the last row).}
{\centering
\begin{tabular}{lcccc}
\hline     
Band     & $m_V - m_D$ & $m_V - m_P$ & $m_V - m_S$ & $m_V - m^c$   \\
\hline
g & 0.057 & 0.024 & 0.050 & 0.043 \\
r & 0.036  & 0.007 & 0.014 & 0.016 \\
i & -0.004  & -0.016 & -0.017 & -0.014 \\ 
\hline
Weights & 20/105 & 35/105 & 50/105 & $-$ \\
\hline
\end{tabular}
 }
\label{tab:offsets} 
\end{table*}

\begin{figure*}
\begin{center} 
\includegraphics[width=0.9\columnwidth]{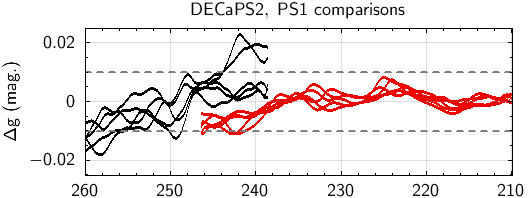}
\includegraphics[width=0.9\columnwidth]{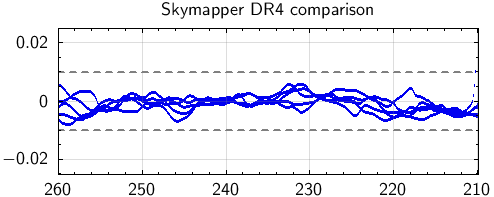}
\includegraphics[width=0.9\columnwidth]{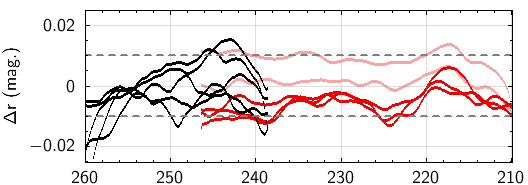}
\includegraphics[width=0.9\columnwidth]{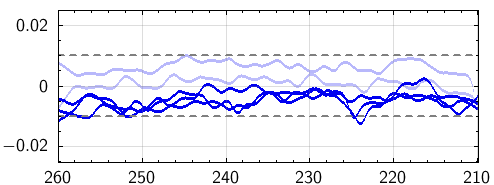}
\includegraphics[width=0.9\columnwidth]{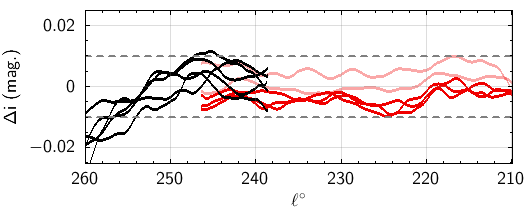}
\includegraphics[width=0.9
\columnwidth]{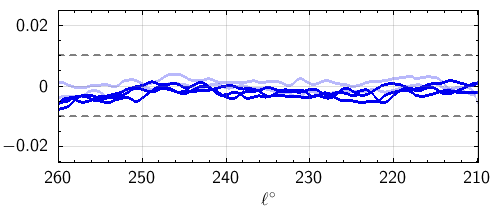}
\caption{Running medians of the differences, $\Delta m$, between Q3-bright listed and colour-equation predicted $gri$ magnitudes along five $2^{\circ}$-wide latitude strips.  The left hand panels shows the differences obtained using the DECaPS-2 (black) and PS1 (red) colour equations.  The right hand panels contain the Skymapper DR4 comparisons (blue).  Top row: $g$, middle: $r$, bottom: $i$.  If all four surveys were completely uniform all differences would be zero. Most of the variation is contained within $\pm$0.01 (drawn as grey horizontal dashed lines). In $r$, a small gradient with increasing $b$ is clearly present: the running median lines, $+3^{\circ} < b < +5^{\circ}$ and $+1^{\circ} < b < +3^{\circ}$, are shown in lighter shades of red (PS1) and blue (Skymapper DR4).  The comparisons with DECaPS-2 indicate a longitude gradient in all bands, not seen in the Skymapper DR4 comparisons. 
}
\label{fig:uniformity}
\end{center}
\end{figure*}

The sought-after zero point correction is the median offset for each band, determined from the full set of colour-corrected magnitude differences.  It is very slightly different from the constant in the linear fit, which is a much closer measure of the mean difference.  The medians are set out in Table~\ref{tab:offsets}.  The DECaPS-2 offsets consistently exceed those obtained via the PS1 comparison: in $g$ and $r$ the difference is $\sim$0.03 mag. This is evidence that these two surveys, whilst aligned off the plane, lose some alignment in the plane. The independently calibrated Skymapper DR4, on the other hand agrees with DECaPS-2 in signalling a $\sim$0.05 mag. offset in $g$, while yielding similar offsets to PS1 in $r$ and $i$. 

The lesson from these comparisons is that there is not a clearly-converged set of $gri$  zeropoint adjustments that we can apply to Q3-bright.  What should we have expected? The pipeline calibration of VPHAS+ magnitudes in the AB system made use of the APASS DR9 all-sky catalogue that was available as a reference set from early on in data taking \citep[see section 2.3 in][]{Drew2014}.  We have followed this up with a homogenization procedure, using Gaia EDR3 photometry, that seeks to iron out non-uniformity across the plane -- it does not provide a global correction to any error that pre-exists in the average APASS DR9 zeropoints.  \cite{Tonry2018} have argued that there are non-negligible discrepancies between the zeropoints of the DR9 release of APASS and PS1 (see their Table 1). The largest correction they find is in the $g$ band, and is 0.053$\pm$0.011 mags. Our comparison of DECaPS-2 and Skymapper DR4 magnitudes with VPHAS+ data supports this - although the PS1 comparison itself indicates just 0.024~mag..  For $r$ and $i$ there is less of a difference.  It appears that zeropoint uncertainties of up to $\sim$0.02 mag. are endemic, particularly given that the alignment of PS1 and DECaPS-2 is degraded in the plane to this extent.   

So far we have considered offsets as they average over large sky areas spanning significant contrasts in typical extinctions (ranging from one to a few visual magnitudes in our Q3 region).  We can now use the global linear colour equatons to get a better picture of how uniform the Q3-bright photometry is.  We break down the Q3-bright catalogue into latitude strips two degrees wide and look at how the difference between measured source magnitudes and their predicted values, vary with longitude along each strip. In Fig.~\ref{fig:uniformity} the running median, smoothed to match strip width, is plotted for each of the 5 strips in each of the $gri$ filters. The smoothing is introduced to remove finer-scale jitter that can hide the strip-to-strip shifts.

The longitude range in which the comparison is with PS1 ($\ell < 245^{\circ}$) presents greater uniformity (mostly to within $\pm 0.01$ mag) than is seen relative to the DECaPS-2 survey ($\ell > 240^{\circ}$).  In the latter case, a small net downward gradient is present, increasing the amplitude of difference to more nearly $\pm 0.02$~mag.  This is accompanied by some higher-frequency spatial variation.  This behaviour may in part be a response to the rise in typical extinction with increasing longitude.  However we suspect the comparison is betraying some inhomogeneity in DECaPS-2.  Evidence that this is likely comes from the comparisons with Skymapper DR4 (right panels in Fig.~\ref{fig:uniformity}): in the $240^{\circ} < \ell < 260^{\circ}$ range, both gradients and fluctuations are notably lower, 

Fig.~\ref{fig:uniformity} exposes a latitude gradient in the $r$ band, such that Q3-bright $r_r$ and $r_b$ magnitudes (either considered separately, or combined into mean $r$) rise by up to $\sim$ 0.02 mag. between $b < 0^{\circ}$ and $b = +5^{\circ}$,  A more modest gradient may affect the $i$ band as well, but this is less clear as only the PS1 comparison indicates this (and not the Skymapper comparison).

\subsection{Comparisons with synthetic tracks}
\label{sec:tracks}

Next, we establish the offsets that might be required for the H$\alpha$ and $u$-band magnitudes.  To do this, we apply the corrections from Table~\ref{tab:offsets} to the Q3-bright $gri$ magnitudes and compare with synthetic photometry.  The average $gri$ shifts carry through to a reduction of the $(g-r)$ and $(r-i)$ colours by 0.027 and 0.030 mag. respectively. Colours or magnitudes corrected according to this prescription henceforth carry a superscript 'c' suffix (e.g. $r_r^c \equiv r_r - 0.016$ mag.).

\begin{figure}
\begin{center} 
\includegraphics[width=0.85\columnwidth]{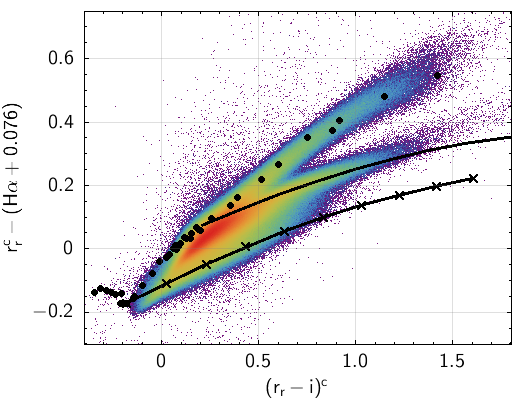}
\includegraphics[width=0.85\columnwidth]{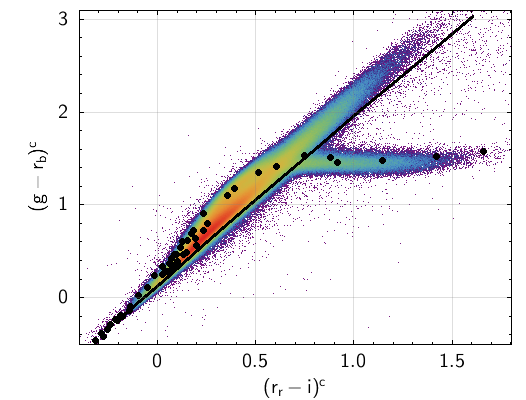}
\includegraphics[width=0.85
\columnwidth]{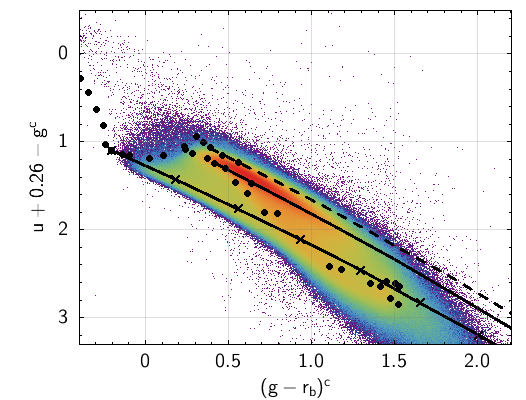}
\caption{Q3-bright colour-colour diagrams, based on corrected $gri$ magnitudes.  Synthetic colours of unreddened dwarfs, with [Fe/H] = 0, and $-0.6$ are superposed as black dots. A2V reddening tracks are included in all panels as black lines. The top panel also includes a K0 III [Fe/H] = 0.0 reddening line (in black). Crosses are placed on the A2V line, at $\Delta A_V = 1$ intervals, in the top and bottom panels.  In the bottom panel, G0V reddening lines are shown for $[Fe/H] = 0.0$ (solid black line) and for $[Fe/H] = -0.6$ (dashed). In all panels, the stellar data are coloured according to a logarithmic density scale (red, high). 
}
\label{fig:vega_ccds}
\end{center}
\end{figure}

Synthetic magnitudes have been computed for main sequence (MS) and giant stars using the measured VST/OmegaCam band transmission profiles and
flux-calibrated library spectra (from the INGS collection\footnote{The spectra, curated by A. Pickles, can be obtained from  https://www.stellarspectralfluxlibrary.com/}). 
Since metallicity in the thin disk, beyond the Solar Circle, falls from [Fe/H] $\sim$ -0.1, down to $\sim -0.6$ \citep{Lian2023}, we deploy synthetic photometry for B to mid-M stars with [Fe/H] = 0 and also for [Fe/H] = -0.6 dex FGK stars. Lowered line-blanketing at reduced metallicity has a noticeable impact on $u-g$ colour that can not be overlooked.  In all reddening lines plotted, the $R_V = 3.1$ extinction law due to \cite{Fitzpatrick1999} is adopted.

Fig.~\ref{fig:vega_ccds} shows the three colour-colour diagrams, obtained from the entire third-quadrant region up to the $\ell = 260^{\circ}$ limit.
In all 3 panels, synthetic MS colours and the A2V-star reddening line are superposed.  These are supplemented by a solar-metallicity K0 III reddening line in the $riH\alpha$ diagram, and by [Fe/H] = 0.0,-0.6 reddening lines for G0V, in the $ugr$ plane.  The relative positioning of the observed and synthetic data is improved in the $gri$ diagram by the small zeropoint corrections applied. After correction, the A2V reddening line is better-positioned with respect to the straight edge of the main stellar locus. In contrast, the colour-colour planes involving the $u$ and narrowband $H\alpha$ filters (bottom and top panels in Fig.~\ref{fig:vega_ccds}) do not align without significant adjustment of their vertical axis scales.  
 
To achieve alignment in the $riH\alpha$ plane, leaving 
$r_r^c$ and $i^c$ alone, the $H\alpha$ magnitudes need to be raised by between 0.06--0.00 mag. This adjustment places centres the main locus between the MS stars and the A2V reddening line.  At these red wavelengths, metallicity has negligible impact. 

Metallicity does matter in the case of the $ugr$ colour-colour diagram (bottom panel, Fig.~\ref{fig:vega_ccds}).  Here, both the $[Fe/H] = 0.0$ and $[Fe/H] = -0.6$ reddening lines for G0V stars are shown (the latter as a dashed line).  This spectral type sits at the turning point of the MS track, and its reddening line should run along the upper edge of the main stellar locus in the ideal case of zero photometric error: reduced metallicity lowers $u-g$, pushing the upper edge further up (on the conventionally inverted vertical axis).  Taking this behaviour into account, along with the placement of the $[Fe/H] = 0.0$ A2V reddening line that marks the lower edge of the main locus, we see that a positive shift of 0.23--0.28~mag. is needed for alignment along the $u - g$ axis.  Given that $g^c$ already carries its correction, the extra shift belongs entirely to the $u$ magnitudes.  That there should be a substantial $u$ correction is no surprise as this band has been recognised before as having an especially uncertain calibration \citep{Drew2014,MMS2017}. 

\begin{table}
\caption{
Approximate shifts to add to VPHAS+ $H\alpha$ and $u$ magnitudes according to the set of zero-point corrections adopted for $g$, $r$ and $i$.  
}
{\centering
\begin{tabular}{lcc}
\hline       
$gri$ correction set     & H$\alpha$ shift & $u$ shift \\ 
\hline     
Mean correction & 0.076 & 0.26 \\
DECaPS-2 & 0.062 & 0.24 \\
PS1 & 0.087 & 0.28 \\
Skymapper DR4 & 0.080 & 0.23  \\ 
\hline
\end{tabular}
 }
\label{tab:ha_and_u_fixes} 
\end{table}
 
 The exact shifts needed to bring the H$\alpha$ and $u$ magnitudes onto a photometric scale that makes sense of the colour-colour diagrams depend on the $gri$ corrections applied. Table~\ref{tab:ha_and_u_fixes} pairs these shifts with the adopted set of $gri$ zeropoint corrections.  The by-eye alignment has a subjective element and there will be uncertainties in the region of $\sim$0.01 mag. in $H\alpha$, and more in $u$ ($\sim 0.02$ mag.). 

\subsection{VPHAS+ and IGAPS photometry compared}
\label{sec:igaps}

Whilst the overarching survey concept of including H$\alpha$ narrow band is common to VPHAS+ in the south and IGAPS, the merged IPHAS and UVEX surveys in the north, there are significant practical differences.  Important among them are the differences in the transmission profiles of the filters used in the VST and INT cameras, and the different methods of pipeline photometric calibration (the IGAPS re-calibration used PS1, and not APASS DR9).  

Another difference concerns observing conditions.  The overlap zone, straddling the celestial equator at 6--7 hours in RA, has to be observed from the ground in either the northern winter, or the southern summer.  This places VPHAS+ at an advantage regarding data quality.  Indeed, the IGAPS data in the 3rd quadrant zone of overlap were a struggle to obtain, accumulating slowly over a decade of winter seasons. As a result, the quality of the IGAPS catalogue in this part of the sky is more uneven than in much of the rest of the northern plane.  This is particularly an issue for the blue filter sets, where there is also a problem of some missing fields at positive Galactic latitude.  This leads to more scatter in the relative colour equations.

\begin{figure}
\begin{center} 
\includegraphics[width=0.85\columnwidth]{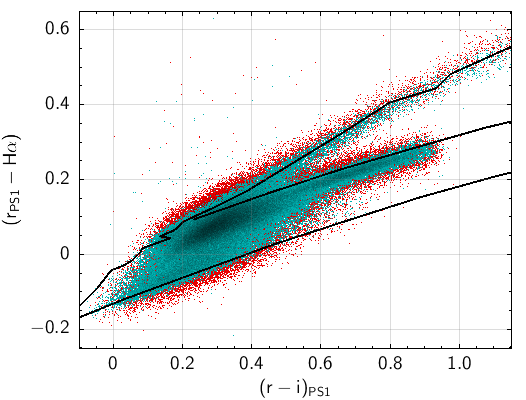}
\includegraphics[width=0.85
\columnwidth]{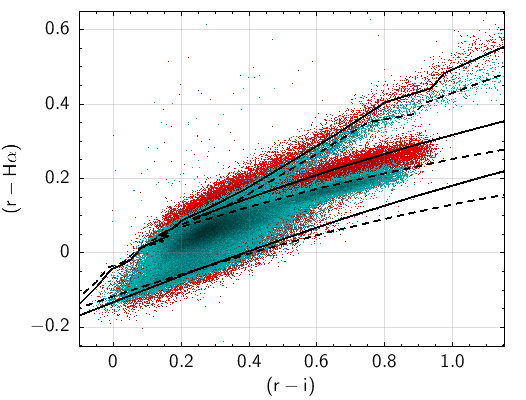}
\caption{VPHAS+ and IGAPS $riH\alpha$ diagrams compared in the overlap zone ($\ell < 215^{\circ}$). 
Top panel: the VPHAS+ $r$ and $i$ magnitudes have been converted to the PS1 system using equations (2), while VPHAS+ $H\alpha$ mags have been increased by +0.095 mag.  VPHAS+ data are in cyan, and sit on top of the IGAPS data in red. The IGAPS $H\alpha$ magnitudes have been decreased by 0.019 mag to improve the alignment. (see text).  Bottom panel: the IGAPS data are as in the top panel, while here the VPHAS+ $r$, $i$ magnitudes are back in the native VST system ($H\alpha$ magnitudes are as in the top panel.  In both panels, the solid black lines are synthetic tracks in the PS1 system (A2V and K0III reddening lines, unreddened MS).  The dashed lines in the bottom panel are the same tracks, but in the native VST system.}
\label{fig:vphas_igaps}
\end{center}
\end{figure}

We cross-match VPHAS$+$ with IGAPS in the overlap spanning the Galactic longitude range from $\ell \sim208^{\circ}$ to $\sim215^{\circ}$ (see Fig.~\ref{fig:density_map}).
The maximum source angular distance permitted in the match is 0.5~arcsec. The error limit imposed on $g$, $r$, $i$ and H$\alpha$ in both surveys is $<0.01$ mag., creating a sample of 106k stars in common, with few fainter than $r = 17$.  For this comparison, the PS1-derived corrections to the VPHAS+ data (2nd column, Table~\ref{tab:offsets}) are appropriate, since PS1 was used as the external reference in the calibration of IGAPS.  As VPHAS+ H$\alpha$ photometry is no more likely to be perfectly uniform than it is in $r$, the $H\alpha$ correction has been redetermined in the IGAPS overlap zone, and we find it to be 0.095, (rather than 0.087 mag. the relevant overall Q3-bright average listed in Table~\ref{tab:ha_and_u_fixes}). 

The best fit linear colour equations are:
\begin{align}
   i^c_V - i_I \ & = &-0.0055 + 0.0026 (g - i)_I & \ \ \ [0.023] \nonumber\\
   r^c_V - \bar{r}_I \ & = &-0.0063 - 0.0298 (g - i)_I & \ \ \ [0.018] \nonumber\\
   g^c_V - g_I \ &=&0.0034 + 0.0836 (g - i)_I &  \ \ \ [0.034] \nonumber\\
   H\alpha^c
   _V - H\alpha_I \ &=&-0.0273 - 0.0007 (g - i)_I & \ \ \ [0.032] 
\end{align}
in which the subscript $I$ indicates IGAPS magnitude, and the number in square brackets at the end of each equation is the fit standard deviation.  We leave $u$ out of this exercise because the IGAPS magnitude ($U_{RGO}$) is uncalibrated and expressed only in the Vega system. 
In the broad bands the colour equations conform very well to the expectation that the zero points fed into the fitting are compliant with PS1: the mean offsets (the constants in equations (2)) between VPHAS+ and IGAPS are all a few millimagnitudes, and the colour gradients hardly differ from the VPHAS+ -- PS1 cross-match equivalents ($c_1$) in Table~\ref{tab:colour_eqns}.  

There is essentially no colour dependence in the relation between the two H$\alpha$ magnitudes also. The narrow-band filters involved greatly limit confounding in-band effective-wavelength shifts, even if the FWHM of the band passes are not quite identical. However there is an offset between the VPHAS+ and IGAPS $H\alpha$ magnitudes of $-$0.027 mag.  It is likely to be due mainly to a drift of the IGAPS H$\alpha$ photometric scale (just as there is drift in VPHAS+ $H\alpha$ magnitudes).  We infer this using a comparison of the VPHAS+ and IGAPS $riH\alpha$ planes (top panel, Fig.~\ref{fig:vphas_igaps}), after fully converting VPHAS+ $r$, $i$ magnitudes into the PS1 system via equations (2). To align the $riH\alpha$ main stellar loci, placing both datasets credibly with respect to synthetic tracks in the PS1 AB system, the IGAPS H$\alpha$ magnitudes need to be reduced by 0.019 mag.  The remaining gap between the two H$\alpha$ scales is, then, less than 0.01 mag. -- a similar residual to those associated with the broad bands.   The greater vertical breadth in the IGAPS main locus probably betrays greater weather impacts on the data.

Whilst the respective $riH\alpha$ planes closely resemble each other, as they should, with broadband magnitudes expressed in the same photometric system, the same expectation does not apply when the broadband systems are different.  We illustrate this in the lower panel of Fig.~\ref{fig:vphas_igaps}, showing a comparison in which the VPHAS+ data are back in the native VST system (with PS1 and $H\alpha$ zeropoint corrections still in place). The main stellar loci of the two datasets are no longer congruent.  The clearest expression of this is the difference in the run of the giant extension, located below the red end of the MS: in VPHAS+, its gradient is shallower than in IGAPS/PS1, and the extension as a whole sits at lower $r - H\alpha$.  These differences are echoed in the synthetic tracks for the two systems overplotted in the figure.  The two $riH\alpha$ planes are not directly interchangeable.

We do not provide a figure comparing the $gri$ planes of the two surveys, as it is enough to note that -- as IGAPS magnitudes are expressed in the PS1 system -- its $gri$ plane closely resembles the case of PS1 in Fig.~\ref{fig:gri_diagrams}. This means the $(g - r)$ dynamic range in IGAPS is as compressed as in PS1, and more than it is in VPHAS+.

\section{From inter-survey comparisons, to Q3-Bright applications}
\label{sec:ABtoVega}

\begin{table}
\caption{Shifts from AB magnitudes to the Vega system, as determined from synthetic photometry.  The numerical correction given should be added to AB magnitudes to obtain the Vega system equivalents.}
{\centering
\begin{tabular}{lr}
\hline
Filter/colour & Correction\\
\hline 
   $i$ & $-$0.349\\
   NB659/$H\alpha$ & $-$0.323 \\
   $r$ & $-$0.137 \\
   $g$ & 0.135 \\
   $u$ & $-$0.878 \\
   $r-i$ & 0.212 \\
   $r-H\alpha$ & 0.163 \\
   $g-r$ & 0.272 \\
   $u-g$ & $-$1.010 \\
\hline  
\end{tabular}
 }
\label{tab:conversions} 
\end{table}

\begin{figure*}
   \centering
   \includegraphics[width=1.95\columnwidth]{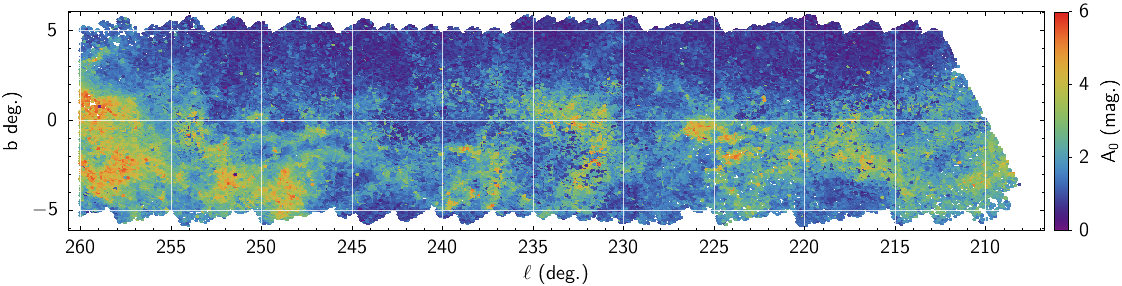}   \includegraphics[width=1.95\columnwidth]{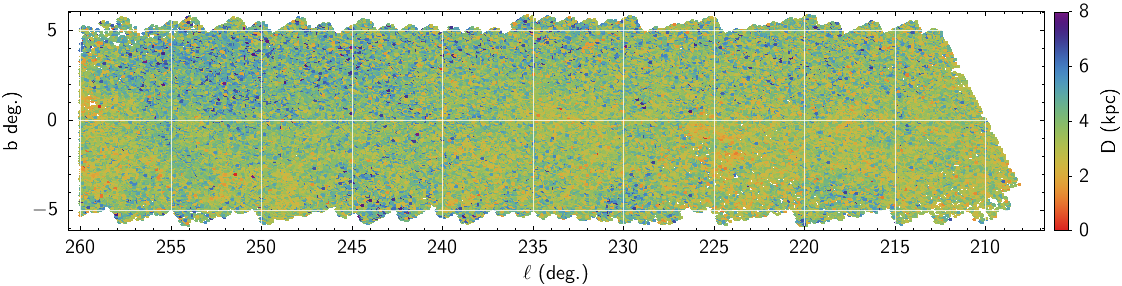}
   \includegraphics[width=1.95\columnwidth]{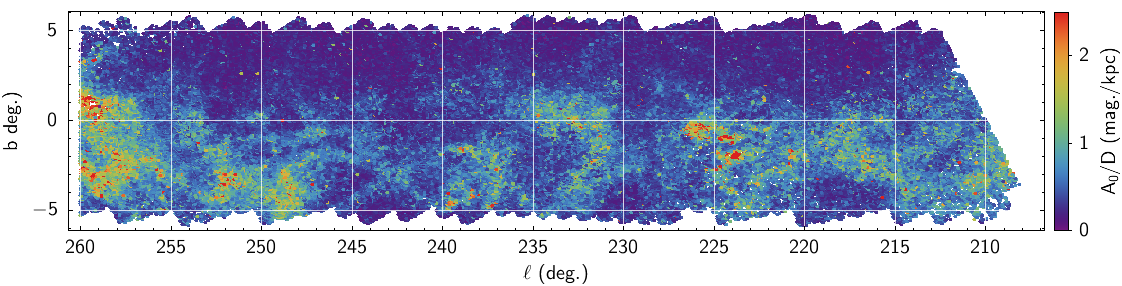}
\caption{Maps of the region constructed using A-star photometry. Upper panel: extinction per star as determined by $r- i$ excess, under the assumption of a mean A3V spectral type.  Middle panel: distance to each star from the Gaia DR3 parallax.  Lower panel: extinction, A$_0$, per kpc - combining the information of the upper and middle panels.  This reveals that for much of the region, extinction accumulates with distance relatively slowly at rates of $\sim$0.5 mag/kpc or less. 
    }
\label{fig:Astar_maps}
\end{figure*}

Our comparisons between VPHAS+ and comparable overlapping surveys, in the third quadrant, are now complete.  Minor zeropoint offsets appear to be appropriate for the $g$, $r$ and $i$ bands, while the $H\alpha$ and $u$ corrections are larger: in the remaining sections of the paper we will use the area-weighted mean corrections from Tables~\ref{tab:offsets} and \ref{tab:ha_and_u_fixes}.  Non-uniformity of the photometry is generally limited to within $\pm0.01$ mags.

The goals now shift to (i) some demonstrations of the application of the Q3-bright catalogue, that reveal the character of the survey and the third quadrant (ii) sanity checks on, in particular, the large $u$ band correction identified in Section~\ref{sec:tracks}.  Some of this will involve calling on existing lists of specific classes of star.  As much published stellar photometry uses the Vega system, conversion from AB to Vega magnitudes becomes necessary.  In Table~\ref{tab:conversions} we provide the numerical data needed for the conversions.  These have been determined via synthetic photometry, in which the OmegaCam filter-plus-detector transmission profiles are folded through CalSpec standard-star spectra (\cite{Bohlin2014}, \cite{Bohlin2020}, \cite{Bohlin2022}) and integrated: essentially, the shift per band is the difference between the AB and Vega-system magnitudes of Vega.  For each band, We synthesize the former, and adopt 0.023 mag. for the latter in all bands from \cite{Bohlin2007}. A subscript 'V' will be appended to symbols representing magnitudes and colours in the Vega system in the folowing sections.

\section{A-star extinction mapping}
\label{sec:astars}

\begin{figure}
   \centering
   \includegraphics[width=0.8\columnwidth]{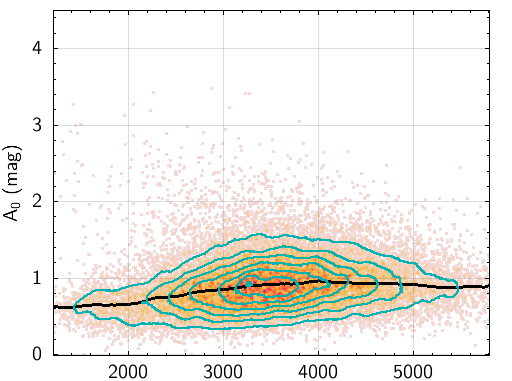} 
   \includegraphics[width=0.8\columnwidth]{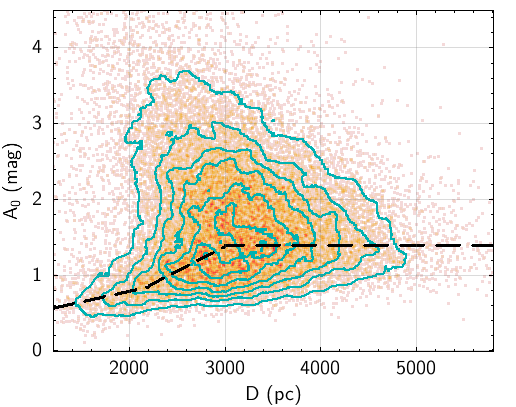}
\caption{Growth in $A_0$ with distance for two $210^{\circ} \lesssim \ell < 240^{\circ}$ strips at contrasting latitudes.  Only {\sc rplx} $> 4$ A-star candidates are included for better discrimination. The upper panel, shows the distribution for stars in the latitude range $1^{\circ} < b < 4^{\circ}$.  The lower panel shows the same for the latitude range $-3.5^{\circ} < b < 0^{\circ}$.    The solid black line in the upper panel is the smoothed median of the positive-latitude data points, while the dashed black line in the lower is a possible interpretation of trend -- inspired by the lower spurs forming the base of the 'tricorn' of contours. 
}
\label{fig:A0_trend}
\end{figure}

The first application we demonstrate is the selection and exploitation of near-MS A stars for extinction mapping the third-quadrant disc.  The selection of these intrinsically relatively-luminous stars is straightforward in concept: they present the strongest H$\alpha$ absorption along the MS and so can be found along the lower edge of the main stellar locus in the $riH\alpha$ diagram.  The detail of how this is done here is presented in Appendix~\ref{sec:Astar_selection},  As distance is an essential parameter in extinction mapping, cross-matching of the A-star selection with Gaia to access parallaxes is appropriate, along with the imposition of a minimum quality requirement on those parallaxes. To maintain good source density and eliminate the lowest quality astrometry, the ratio of measured parallax to error ({\sc rplx}, is required to exceed 2.  The sample of photometrically-selected A stars created this way contains 225k stars.  

Distances ($D$) are inferred from Gaia DR3 parallaxes using the EDSD method introduced by \cite{Luri2018}, with a length scale of 1500 pc and a zero-point offset of 17 $\mu$arcsec.  Monochromatic extinctions at 5500~\AA , $A_0$, are calculated from $(r_r - i)^c_V
$ (converted into Vega magnitudes), by interpolating data from the main sequence colour tables in \cite{Drew2014}.  A median spectral type of A3V, and an $R_V = 3.1$ extinction law are adopted. 

Maps of the region relating to extinction, distance and the ratio between them are presented in Fig.~\ref{fig:Astar_maps}.  The two-dimensional extinction map in the top panel of this figure shows how the dust distribution is strongly asymmetric, being greater at Galactic latitudes below $b \sim +1$.   At higher latitudes, $A_0$ exceeds 2 magnitudes only rarely, while below $b = 0^{\circ}$, 4--5 magnitudes is reached in some patches.  Since these stars are among the more distant, and therefore reddened, of all of the stars in the VPHAS+ catalogue, more frequent, cooler, lower-luminosity stars will be less reddened.

With the incidence of larger dust columns higher at and below the Galactic equator, the longest A-star sightlines are more common at positive Galactic latitudes (middle panel of Fig.~\ref{fig:Astar_maps}).  The bluer cast of the distance map there indicates distances $>5$~kpc are not unusual.  The median of the distances across the whole map is 3.72 kpc.  Below the Galactic equator, distances tend to be less than this, giving the map a $\sim$yellow cast.  But only 86 stars in the entire sample have parallaxes consistent with distances below 1 kpc.  Some of these are hot subdwarfs that have crept into the selection at the low-extinction end.

The third quadrant, being outside the Solar Circle, should show lower overall concentrations of dust.  This is borne out by the map of the rate of extinction increase with distance (bottom panel of Fig.~\ref{fig:Astar_maps}).  Over much of the region $A_0/D$ is well under 1 mag kpc$^{-1}$, with the median being 0.39 mag kpc$^{-1}$.  This is low for the Galactic plane.  For example, in the 4th quadrant, sightlines intersecting the near and far Carina Arm commonly exhibit rates of between 0.8 and 2 mag kpc$^{-1}$ \citep[see Fig. 19 in][]{MMS2017}.  At still higher longitudes higher rates will prevail.

The continuation of the Perseus and Local Arms should be the main features, at distances of 3--6 kpc and 1--2 kpc respectively \citep[see e.g.][]{Reid2019}.  The obvious dust clouds apparent around $\ell \sim 225^{\circ}, b \sim -1^{\circ}$ are associated with CMa OB1, in the Local Arm. This has been placed at $D = 1209 \pm4 \pm 60$ by \cite{Zucker2019}.  Nearly all the VPHAS+ A stars in this region are behind it.  Towards $\ell = 260^{\circ}$, the beginning of the Vela Molecular Ridge -- described by \cite{Murphy1991} as marking one end of the Local Arm -- is apparent as the patches of highest extinction in the map.  Part of 'Cloud D', in the nomenclature of \cite{Murphy1991}, sits close to the Galactic equator at a distance of $\sim$ 1 kpc, with Vela R1 at lower latitudes, and under 500 pc away.  In between ($230^{\circ} \lesssim \ell \lesssim 255^{\circ}$), either on or below the Galactic equator, other arcs of raised extinction are apparent with $A_0 \sim 1$, accumulating. 

The presence of the Perseus Arm is not immediately obvious: \cite{Reid2019} links three molecular clouds to the arm in this part of the plane -- G229.57+0.15, G236.81+1.98 and G240.31+0.07 -- and none of them leaves a noticeable imprint on the maps in Fig.~\ref{fig:Astar_maps}.  However, if the third dimension, distance, is brought into play, more insight can be achieved.   In Fig.~\ref{fig:A0_trend}, the distance distributions of A-star candidate extinctions,  extracted from two constant-latitude strips ($210^{\circ} \lesssim \ell < 240^{\circ}$) above and below the Galactic equator, are compared.  

The positive latitude strip is 'quiet' in that there is a slow rise in extinction with distance, plateau-ing at almost $A_0 = 1$ mag from the distance of $\sim$4 kpc.  Below $\sim$1.5~kpc, there is little to no information as there are few stars.  We can conclude that the Local Arm is responsible for the first $\sim$0.6 mag of extinction.  

At negative latitudes the picture is more complex. The peak density of points lies at $\sim$3 kpc, and $A_0 \sim 1.4$~mag.  There is a gradual decline in source density to larger $A_0$ that will be associated with the darker clouds of extinction (e.g. near CMa OB1) visible in the 2D map (Fig.~\ref{fig:Astar_maps}, top panel).  Where the contours crowd near the base of the contour triangle, a minimum general ISM trend (analogous to that at positive latitudes) suggests itself. This is marked with a dashed line in Fig.~\ref{fig:A0_trend}. The difference is that it plateaus at higher extinction after a sharper rise from $\sim$2 kpc, as distance increases to the $\sim$3~kpc contour peak.  This extra increase of under 1 magnitude of extinction may occur in the Perseus Arm. Like the Local Arm, it is offset to negative Galactic latitudes.

\section{The  properties of selected object classes: cross-matches and colour-colour diagrams}
\label{sec:SIMBAD}

The object classes we examine next are (i) sub-luminous stars, (ii) red giants, and (iii) young stellar objects. To assist in this we cross-match VPHAS+ with SIMBAD (and other catalogues) in order to gather large samples of more specialist object types from across the whole region.   
Setting a maximum angular separation of 1 arcsec, in the SIMBAD cross-match with Q3-bright, we obtain a list of 88,211 objects.  For 98\% of the cross-matches, the angular offset is actually no more than $\sim$0.25~arcsec.  Raising the limit to 1~arcsec has the virtue of bringing into the sample a significant number of higher proper motion objects (e.g. white dwarfs). 

In this section and in sections~\ref{sec:emstars} and \ref{sec:OBstars}, photometric magnitudes will be presented and discussed in the Vega system.

\subsection{Sub-luminous stars}
\label{sec:WDs}

\begin{figure}
\begin{center}
\includegraphics[width=0.89
\columnwidth]{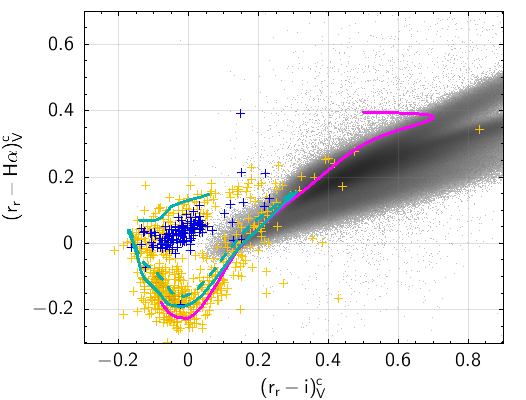}
\includegraphics[width=0.9
\columnwidth]{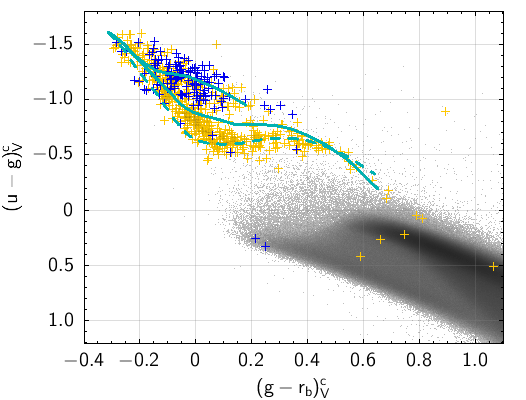}
\caption{
Properties of SIMBAD-cross-matched white dwarfs and hot subdwarfs (Vega magnitudes) in the $riH\alpha$ and $ugr$ colour-colour planes. Top panel: the $riH\alpha$ plane, with Q3-bright sources defining the main stellar locus in grey.  WDs are marked as yellow crosses, hot subdwarfs as blue crosses.  The cyan tracks superposed are theoretical DA and DB tracks from Raddi et al. (2016), for $\log g = 8.0$. The dashed cyan lines show the DA tracks for $\log g = 7.5$.  The DB track, in contrast, is gravity insensitive.  The magenta line is a Bergeron DA track (from Verbeek et al. 2012, computed for the IPHAS filter set), This extends the $T_{{\rm eff}}$ range down to 3 kK.  Bottom panel: the $ugr$ plane.  Symbols and lines as in the top panel.
}
\label{fig:subluminous} 
\end{center}
\end{figure}

After objects with more doubtful Gaia DR3 astrometry (RUWE $>$ 1.4) have been removed, the numbers of cross-matches in the main sub-luminous classes are 555 white dwarfs, 60 white-dwarf candidates, 128 hot sub-dwarf candidates, and 5 hot sub-dwarfs.  Too few interacting binaries are present to be statistically useful.  We also set aside the candidate white dwarfs as there is no need to consider them in the presence of nearly 10 times as many higher confidence examples.  However all the candidate hot subdwarfs must be retained. 

Of the 555 white dwarfs (WD), all but 21 are discoveries from trawling the Gaia catalogue \citep{GF2019, GF2021}.  In order to find a match with the Q3-bright catalogue, all must be brighter than $r = 19$.  Unsurprisingly for such intrinsically-faint objects, the distribution created by such a cut is very much piled up at the faint end: $\sim70$\% are fainter than $r = 18$.  On sky, the distribution is relatively uniform.  The median distance inferred from the generally large and precise Gaia DR3 astrometric parallaxes is $\sim$240~pc.  83\% of the sample are assigned DA spectral types \citep[mostly from][]{Vincent2024}.  The hot sub-dwarfs, in contrast, spread in magnitude from $r \sim 14$ right down to the faint limit.

The location of these objects in the $riHa$ and $ugr$ colour-colour planes are shown in Fig.~\ref{fig:subluminous}. The WDs are plotted as yellow crosses, while the hot sub-dwarfs are marked in blue.  In both diagrams, most of them sit clear of the main stellar locus (in grey).  

In the $riH\alpha$ plane, the limited number of WD in the main stellar locus, are on or near the cool end of the superposed $\log g = 8.0$ DA theoretical tracks, taken from the tables presented by \cite{Raddi2016} and by \cite{Verbeek2012}.  The DA and DB tracks from \cite{Raddi2016} were obtained by folding \cite{Koester2010} synthetic spectra through VST filter profiles. A representative small amount of reddening has been applied to them ($E(B-V)$ = 0.03, or $A_0 \simeq 0.1$). The DA and DB effective temperature ranges covered are, respectively, 6~kK -- 100~kK, and 10~kK -- 40~kK. The \cite{Verbeek2012} Bergeron DA track is shown in Fig.~\ref{fig:subluminous} as well, solely to extend the theoretical track to lower effective temperatures (down to 3~kK -- it goes no hotter than 17~kK).   
For the DA WDs there is clear sensitivity to surface gravity, as illustrated by the difference between the $\log g = 8.0$ and $7.5$ tracks.  In $riH\alpha$, the changing strength of the H$\alpha$ damping wings will be the cause. The DB WD track is gravity-insensitive as DB spectra show little/no hydrogen line absorption. Among the WDs plotted there is a spread that can be due to either or both $\log g$ (for the DAs) and extinction.

The selection of WD from the $ugr$ plane (bottom panel, Fig.~\ref{fig:subluminous}) is easier, because of the better separation achieved in the $(u - g)$ colour.  Again, there is sensitivity to surface gravity among the DA WDs, but at the same time, extinction shifts can be expected to be more pronounced.  The positioning of the WDs in $ugr$ supports the need to substantially increase the $u$ magnitudes as indicated in Table~\ref{tab:ha_and_u_fixes}. But it is not very sensitive. The averaged offset of $+0.26$ mag. does a reasonable job of keeping the observed WD colours close to the DA and DB theoretical tracks. Certainly, as the offset is lowered, more and more WDs become improbably blue in $(u - g)$.  Raising the offset to as much as e.g +0.3 is a 'small' shift in context, and would not be unduly problematic -- but a further increase would begin to imply surface gravities for the bulk of the DA population that are too low.     

Among the WDs and also the cross-matched hot subdwarfs there are sources made appreciably redder by binarity. These are more likely to merge in with the main stellar locus in both colour-colour planes. Most of the hot sub-dwarfs plotted appear in the catalogue constructed by \cite{Geier2019}, using Gaia DR2 data.  These authors offer binarity and the presence of cool companions as the explanation for such objects \citep[see also][]{Dawson2024},  The majority of the hot sub-dwarfs locate near or mingle in with the DB white dwarfs, consistent with their high effective temperatures ($> 20$~kK).  In the $riH\alpha$ diagram, their displacement below the DB WD track signals the presence of some H$\alpha$ absorption.

\subsection{Luminous red stars}
\label{sec:LPV}

\begin{figure}
\begin{center}
\includegraphics[width=0.95
\columnwidth]{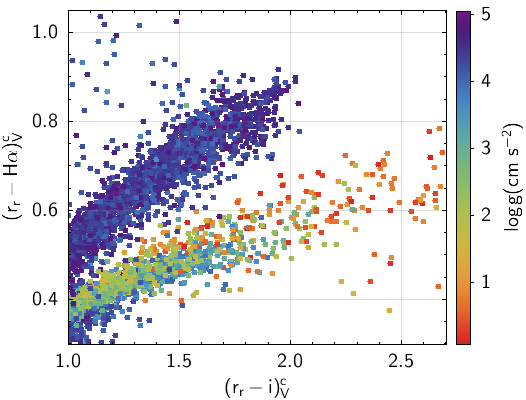}
\caption{Gaia DR3 surface gravities of red stars with SIMBAD counterparts, as they distribute in the $riH\alpha$ colour-colour plane (Vega magnitudes). 
}
\label{fig:red_parameters}
\end{center}
\end{figure}

\begin{figure*}
\begin{center}
\includegraphics[width=0.95\columnwidth]{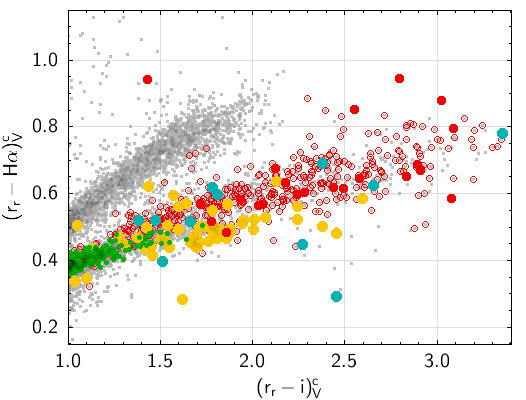}
\includegraphics[width=0.95\columnwidth]{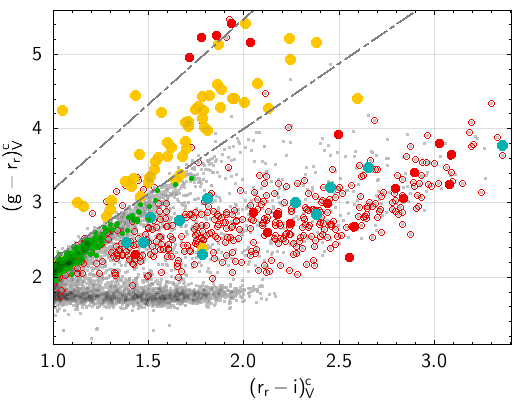}
\caption{Known LPVs and other luminous red stars in the $riH\alpha$ and $gri$ colour-colour planes.  In both panels, all the SIMBAD-cross-matched stars with $(r_r - i)^c_v > 1.0$ are shown in grey.  Red open circles mark objects typed as long-period variables (LPV) by SIMBAD.  Green picks out stars classified as RGB stars.  The larger filled red circles are Mira stars, with carbon stars shown in yellow and S stars in cyan. The grey dash-dotted lines in the right hand panel are approximate boundary indicating the region that might be used to select carbon stars. 
}
\label{fig:LPVs_plus}
\end{center}
\end{figure*}

A feature of the $riH\alpha$ diagram that was noticed and explored in IPHAS data \citep{Wright2008} was an especially-marked separation of S-type stars from the main run of extreme red giants.  For the carbon-rich C-type stars, there were hints of a similar, but weaker effect.  Is the same seen in VPHAS+ data?  We address this question after recalling the general properties of the reddest stars present in the survey region.

For present purposes, a red star satisfies $(r_r - i)^c_V > 1.0$ (Vega, equivalent to a cut at $\sim$0.8 in the AB system).  On the main sequence, unreddened stars satisfying this criterion would be $\sim$M2V or later in type.  The red stars, as defined in this way, that have SIMBAD counterparts, are cross-matched with the Gaia DR3 catalogue.  The resultant list contains 5364 objects.  Of these, 3704 are accompanied by DR3 stellar parameters that have been estimated, aided by Gaia prism data.  

Fig.~\ref{fig:red_parameters} presents the $riH\alpha$ colour-colour diagram for these objects, with the datapoints shaded according to their estimated surface gravities.
The two spurs in the diagram cleanly separate into higher-$\log g$ cool dwarfs at higher $(r_r - H\alpha)$ and a mix of stars dominated by lower surface gravity giants and more extreme objects in the lower spur.  However the intrusion of some blue-shaded data points in the lower spur betrays the presence of some hotter, significantly reddened near main sequence stars.  These objects, along  with cooler first-ascent giants, are evident in the lower spur out to $(r_r - i)^c_v \sim 1.7$, with cool, more luminous, low surface gravity AGB stars taking over at the reddest colours.  

Now, the focus narrows to just those red objects assigned within SIMBAD to the main groups of extreme red giants -- the long period variables (274 stars), Mira variables (20), C and S stars (39 and 9 stars respectively -- 7 of the S stars are 'candidates').  Fig.~\ref{fig:LPVs_plus} shows where these objects fall in both the $gri$ and $riH\alpha$ colour-colour diagrams.  In the IPHAS $riH\alpha$ diagram, both C and S stars tended to exhibit lower $(r - H\alpha)$ than their more O-rich counterparts, represented in the figure by the confirmed long-period variables (LPVs).  There is some sign of this effect, but it is not very strong as there are clearly examples of both C and S stars mixed in with the LPVs (along with some well below the main LPV trend).  In part, this may be a consequence of the H$\alpha$ filter differences between the two surveys, with the VPHAS+ filter being redder and a little wider.  Better statistics are certainly needed for S-type stars.

\begin{figure*}
\begin{center}
\includegraphics[width=0.9\columnwidth]{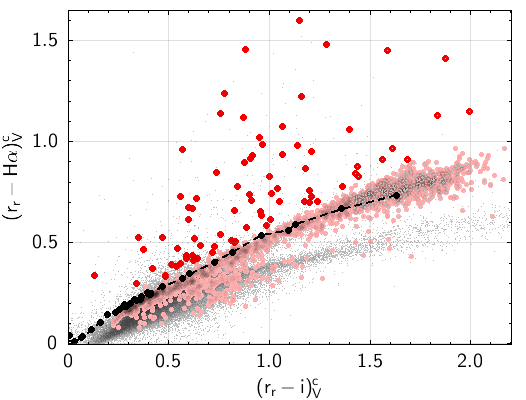}
\includegraphics[width=0.9
\columnwidth]{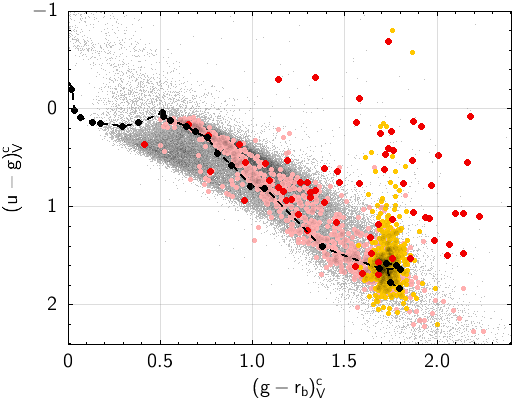}
\caption{SIMBAD 'YSOs' in the Vega-system $riH\alpha$ and $ugr$ colour-colour planes.  All the SIMBAD-cross-matched stars provide the grey background of sources, with pink used for stars carrying YSO-related labels, as discussed in the text.  The red data points represent probable emission line stars by virtue of their location in the $riH\alpha$ plane. 
 The black points and connecting dashed lines are the synthetic unreddened main sequence up to mid M type. In the right panel, yellow is used for non-emission YSO-labelled objects redder than $(r_r -i)^c_V = 1.4$ (mid- to late-M dwarfs); as plotted, these form a bottom heavy vertical sequence.
}
\label{fig:SIMBAD_YSO_riha}
\end{center}
\end{figure*}

Interestingly, the $gri$ colour-colour plane provides much better discrimination between carbon stars and the LPVs (right-hand panel of Fig.~\ref{fig:LPVs_plus}). Aside from a small number of outliers, the SIMBAD cross-matched set of carbon stars can be seen to follow a clear trend lying on the opposite (upper) side of the RGB-dominated spur, from the LPVs. None of the small number of S-type stars/candidates separates out. Most of the carbon stars plotted were first identified as such, via their characteristic C$_2$ bands from objective prism spectra: strong Swan-band absorption in the blue/green optical is the origin of these stars' very red $g-r$ and $g-i$ colours.  Photometric selection using Sloan filters has been deployed successfully before now, as described by \cite{Green2013} in his search for dwarf carbon stars.  VPHAS+ data could be exploited in a similar fashion.  Trialling the selection zone indicated by the dash-dotted lines in Fig.~\ref{fig:LPVs_plus} on the Q3-bright catalogue, without the SIMBAD cross-match, we would find a further 50 to 100 candidate carbon stars in the region. Of course, the higher density of such objects will be in the inner Galaxy, and not outside the Solar Circle.  This is a topic to develop with the forthcoming full VPHAS+ catalogue.  

\subsection{Young stellar objects}
\label{sec:YSO}

Over the last decade the definition of 'young stellar object' (YSO) as deployed in SIMBAD, has loosened.  This reflects a change in the way young stars are selected.  Before the loosening, a YSO was generally an accreting object picked up in star-forming regions via their infrared excesses and/or optical line emission.  Now, the YSO-relevant labels in SIMBAD are also assigned to high-probability members of young open clusters and to stars with estimated positions in the Hertzsprung Russell diagram above the main sequence. Many of these have been added since the arrival of comprehensive Gaia selections of open clusters and stars based on the mission's optical photometry and astrometry \citep[e.g.][]{CG2018,Zari2018,Hunt2023}.  

Our SIMBAD cross-match contains 1941 objects carrying the labels YSO, YSO candidate, TTau star, or TTau candidate. Around a half are from the \cite{Zari2018} catalogue. However, why these labels have been applied is not immediately apparent because of the range of literature sources involved.  Nevertheless, the history of research in this part of the sky has led to a spatial distribution favouring the lower Galactic latitudes -- few locate at $b > 2$. How they distribute in the VPHAS+ $riH\alpha$ colour-colour plane is shown in the left panel of Fig.~\ref{fig:SIMBAD_YSO_riha}: this is plainly a lightly-reddened population, with some emission line stars. The Gaia DR3 GSP-Phot parameters for 1547 out of the 1840 stars, indicate for the great majority distances $\lesssim 1$ kpc, and visual extinction $\lesssim 1.5$ mag.  

The proportion of emission line stars among the SIMBAD YSOs and candidates, suggested by Fig.~\ref{fig:SIMBAD_YSO_riha}, is low at $\sim4$ percent.  However, this proportion rises significantly for samples focused on signatures of accretion.  We illustrate this with reference to two studies in the literature, identifying active accretors. The first of these, by \cite{Prisinzano2018}, searched for YSOs in the Vela Molecular Ridge at $\ell > 258^{\circ}$: they examined stars in the region for NIR excess, for H$\alpha$ emission using VPHAS+ data from the ESO archive, and also for X-ray activity.  Of the 857 YSO candidates spread across the ridge, 329 (38.5\%) were photometrically-plausible H$\alpha$ emitters.  NIR excess was detected more often, in 576 (67.5\%) cases. Interestingly, in only 136 stars were both accretion markers present, leaving many linked to just one.   The second study, of CMa~OB1 around $\ell \sim 224^{\circ}$, by \cite{Sewilo2019}, focused on infrared selection and SED fitting: of the 293 YSOs identified, only 55 have optical magnitudes in the right range to cross-match with Q3-bright.  Around half the 55 turn up as candidate emission line stars among the red data points in the left panel of Fig.~\ref{fig:SIMBAD_YSO_riha}.  

A further potential accretion marker, available from VPHAS+ data, is the blue excess created by the hot inner accretion flow onto the young star. This is accessible via the $ugr$ digram (right panel of Fig.~\ref{fig:SIMBAD_YSO_riha}). Indeed, a sizeable fraction of the YSOs with H$\alpha$ emission picked out in the $riH\alpha$ diagram, appear on the blue side of the main stellar locus in $ugr$.  Some do not, which may flag such stars as further on in their PMS evolution (or as interloping post-MS emission line stars).  The M stars that would form a horizontal spur at constant $(g-r_b)$ in the $gri$ diagram, line up vertically in $ugr$ (picked out in yellow in Fig.~\ref{fig:SIMBAD_YSO_riha}).  However, some of the bluest in $(u-g)$, in this spur (mixing in with the $riH\alpha$-confirmed emission-line objects), may be old WD+M binaries, as described by \cite{Smolcic2004}, rather than young stars.

\section{Emission line star selection}
\label{sec:emstars}

\begin{figure}
\begin{center}
\includegraphics[width=0.77\columnwidth]{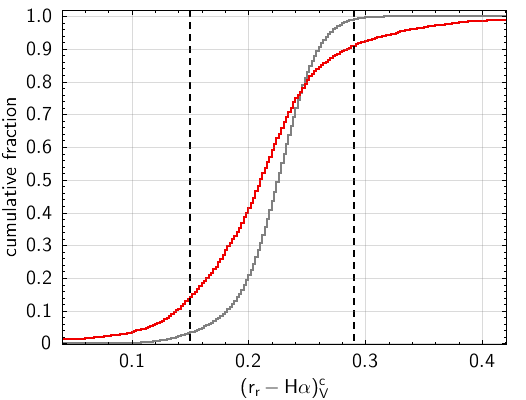}
\caption
{
Comparison of the normalised cumulative function of $(r_r - H\alpha)^c_v$ for eclipsing binaries (red) and Q3-bright as a whole (grey). The included objects (~$\sim 4000$ EB, > one million Q3-bright stars) lie in a strip in the $riH\alpha$ plane, defined by $0.475 < (r_r-i)^c_v < 0.525$. The vertical dashed lines mark the mean crossings within the strip of the A2V reddening line and unreddened solar-metallicity main sequence at $(r_r - H\alpha)^c_V = 0.15$ and $0.29$ respectively -- roughly the bounds to the main stellar locus. 
}
\label{fig:EB}
\end{center}
\end{figure}

An important preliminary to a discussion of emission line star selection is to consider possible contamination.  Not every object, above the unreddened main sequence in the $riH\alpha$ (or $girH\alpha$) diagram, will present line emission in its spectrum. Variability, in particular, can result in misleading colours.  The survey strategy of cycling through all 3 filters per red, or blue, observing block was designed to minimise this.  But, within each OB, there is still some time delay, of the order of 20 to 40 minutes between the acquisition of the different filter images.  In principle, emission line star selection can also be performed using the $girH\alpha$ plane \citep[the analogous diagram using $V - I$ has been exploited by e.g.][]{Sung2004, Sung2008}.  In the case of VPHAS+ data, variability is much more likely to intrude, since the $g$ observation can be months or years apart from the red-filter observations.

Common contaminating variables are the eclipsing binaries (EB) -- particularly short period systems, in or near contact, undergoing nearly continuous light changes.  To explore their impact, we have extracted all the EBs in the region from the catalogue of \cite{Mowlavi2023} based on Gaia DR3, and have cross-matched them with Q3-bright (36,660 objects).  These variable objects spread above and below the main stellar locus. By how much, is illustrated in Fig.~\ref{fig:EB}, where the normalised cumulative $r_r - H\alpha$ distributions for EB and the complete Q3-bright catalogue are compared. The objects included lie within a fixed $(r_r-i)$ strip straddling peak density in the $riH\alpha$ plane.  Almost 10 percent of the EB sit above the unreddened MS.  Indeed, there are around 2000 in the \cite{Mowlavi2023} catalogue that mix into the zone that should belong to the emission line stars. 

\begin{figure}
\begin{center}
\includegraphics[width=0.8\columnwidth]{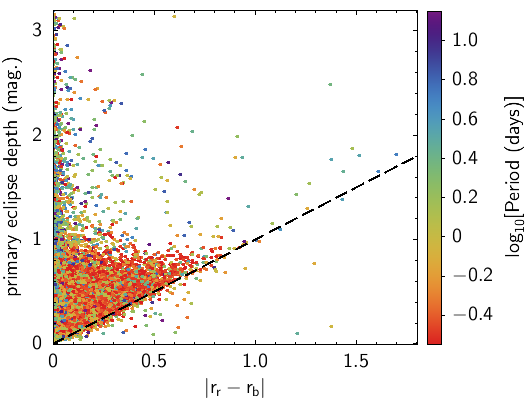}
\caption
{Comparison between VPHAS+ $|r_r - r_b|$ and primary eclipse depth, from Molawi et al (2023) for the cross-matching EBs.   The data points are coloured for estimated orbital period (1/freq, where freq is the quantity in the Gaia DR3 catalogue).  The dashed black line is the equality line. 
}
\label{fig:EB_rdiff}
\end{center}
\end{figure}

The EB are also characterised by a broader distribution in the $r$ magnitude difference, $|r_r - r_b|$. Quantifying this: 90\% of all stars in the strip have $|r_r - r_b| < 0.023$, in contrast to only 43\% of eclipsing binaries.
Fig.~\ref{fig:EB_rdiff} shows how this magnitude difference measures up against the primary eclipse depths, and periods, determined from Gaia DR3 $G$ light curves.  It is indeed the shorter-period systems that dominate the larger $|r_r-r_b|$ measurements.  

\begin{figure}
\begin{center}
\includegraphics[width=0.95\columnwidth]{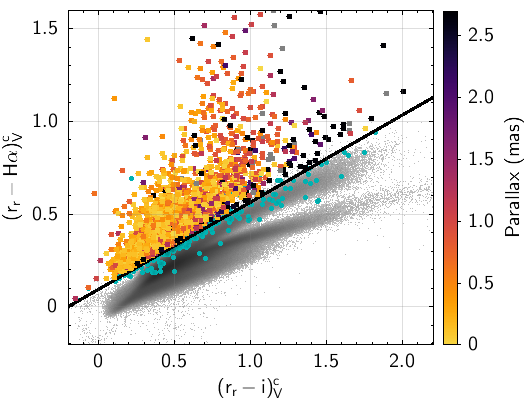}
\includegraphics[width=0.95\columnwidth]{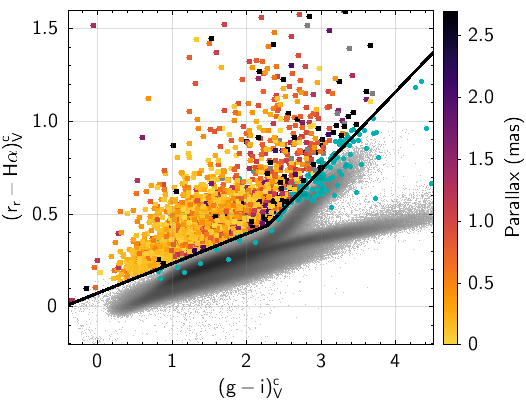}
\caption
{Top panel: Emission line candidate selection in the $riH\alpha$ plane. Lower panel: Emission line selection in the $girH\alpha$ plane.  In both panels the main locus formed from all objects input to selection are shown in grey, while the black lines are the selection lines used.  The latter have been chosen by eye to sit above the main locus where the point density has fallen away to a low level. The data points selected are coloured for Gaia DR3 parallax: M dwarfs, as nearby objects are dark, while distant stars, with bluer photometric colours, shade to yellow.  Cyan is used for candidates only found in the alternative plane: more than twice as many are selected only in $girH\alpha$ (478), than in $riH\alpha$ (200). 
}
\label{fig:emission_selection}
\end{center}
\end{figure}

EBs, mimicking $(r_r - H\alpha)$ colour excess, do not always pair with larger $|r_r - r_b|$. So, screening for large $|r_r - r_b|$ will not remove them all from a search for emission line objects. Worse still, this tactic risks excluding photometric variables that are also emission line stars.  For example, significant YSO variability has been characterized by \cite{Froebrich2022} for the case of YSOs in the $\sigma$~Ori cluster: they found $R$ band variability of up to 0.3 mag on timescales as short as 0.1 days (see their Fig. 7). Partial decontamination of an emission line selection, by removing knowns EB, is the only option. We start with this step, using both the \cite{Mowlavi2023} catalogue (removing objects listed with $G$ eclipse depths $> 0.1$ mag) and SIMBAD (removing objects typed as EB, not listed in the Mowlavi et al catalogue).  This eliminates 35k stars. 

We now appraise the relative performances of $riH\alpha$ versus $girH\alpha$ emission line selection.  In both approaches, we impose the requirement that candidates are offset by at least 10$\sigma$ above a straight line (positioned by eye) that defines the upper boundary of the observed main stellar locus.  Here, $\sigma$ is the error in $(r_r - H\alpha)$, computed from the source's individual band errors, as described by \cite{Monguio20}. To minimise the impact of photometric errors on the appearance of the two colour-colour planes, $g$, $r_r$, $H\alpha$, and $i$ errors are limited to less than 0.01 (leaving 6.69 million stars to select from).  

Using the $riH\alpha$ plane, we obtain a list of 2030 emission candidates, while the selection in $girH\alpha$ returns 2308 objects. These selections are shown in Fig.~\ref{fig:emission_selection}, along with the cut lines used.  For the $girH\alpha$ selection, the boundary has to be made up of two straight-line segments, in order to exclude the M-dwarf spur rising up from the main stellar locus after $(g-i)^c_V \sim 2.3$ mag.

There are 1828 objects in common between the two selections. Most of the 202 stars found only in the $riH\alpha$ selection, are missed in $girH\alpha$ because they overlap the M-dwarf spur (lower panel of Fig.~\ref{fig:emission_selection}).  The candidates found in the $girH\alpha$ selection, but not in $riH\alpha$, number 480. They straddle the $riH\alpha$ cut line, with 138 of them falling below (these are visible in cyan in the top panel of Fig.~\ref{fig:emission_selection}). Photometric variability appears as a factor in this, in that the median $|r_r-r_b|$ for these stars is 0.11 mag., whilst it is only 0.012 mag. for stars above the cut line.  The $girH\alpha$-only candidates above the $riH\alpha$ cut line miss $riH\alpha$ selection because they do not meet the $>10\sigma$ criterion.  In the VPHAS+ context, there is reason to place more trust in the $riH\alpha$, rather than the $girH\alpha$, selection.

\begin{figure}
\begin{center}
\includegraphics[width=0.9\columnwidth]{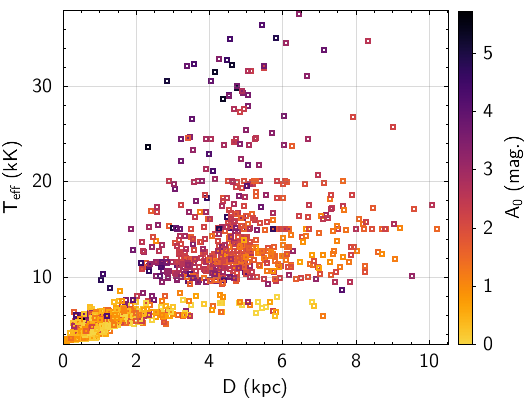}
\caption{
Astrophysical properties of the emission line candidates, common to both the $riH\alpha$ and $girH\alpha$ plane selections, as listed in the Gaia DR3 catalogue. Parameters are available for 1300 stars ($\sim70$\% of the total selection).
}
\label{fig:emission_APs}
\end{center}
\end{figure}

Aided by the astrophysical parameters available in the Gaia DR3 catalogue, for 1300 of the common selection of 1824, we consider some characteristics of the emission line candidates.  Distance, effective temperature and extinction are summarised in Fig.~\ref{fig:emission_APs}. There is a split into two main groupings: a cool closer subset and a generally more distant hotter, more luminous group. Magnitude limits play a role here: lower-luminosity cooler stars are too faint to appear in the distant group, massive stars are too bright for the near group. The current selection scarcely goes fainter than $r \sim 18$, and $r > 13$ is imposed at the bright end.   
 
The emission line candidates favour negative Galactic latitude. The median latitude is $b = -1^{\circ}.40$, to be compared with a median of $+0^{\circ}.37$ for the parent 6.69 million stars. On placing the candidates into near ($D < 2$ kpc) and distant ($3.5 < D < 6.5$ kpc) groups, we find the median Galactic latitudes for them are $b = -0^{\circ}.97$ and $-2^{\circ}.35$ respectively.  Indeed, for the 'far' group, most likely linked with the Perseus and Outer Arms, we find a broad distribution of $z$ offsets below the plane with an inter-quartile range of $-270$ to $-20$ pc.  

\section{OB stars from blue selection}
\label{sec:OBstars}

The leading motivation for the inclusion of the $u$ band observations in VPHAS+ was the possibility to find 'blue excess' objects.  Intrinsically, O and early B stars fall into this class thanks to their high effective temperatures: a variant of the 'Q method', pioneered by \cite{Qmethod}, uses the $ugr$ plane to separate them out from cooler stars. This has already been applied to VPHAS+ data covering the Carina region by \cite{MMS2015, MMS2017}.  Here, we apply this method of OB-star selection in the third Galactic quadrant to see this method of selection at work in this less extinguished environment, and to see how the preliminary $u$ calibration performs.

It was noted in Section~\ref{sec:tracks} that the main stellar locus in the $ugr$ colour-colour plane is broadened by metallicity effects (see the bottom panel in Fig.~\ref{fig:vega_ccds}.  The selection method described by \cite{MMS2015} looks for stars located on the blue side of the B3V reddening line.  In the Carina part of the plane, this is fairly straightforward because the B3V reddening line sits clear of the densely populated main locus. The challenge of the third quadrant is the smearing of the main stellar locus created by large numbers of lower metallicity objects: the [Fe/H] = 0, B3~V reddening line sits almost on top of the [Fe/H] = -1.2 G0~V equivalent.  Significant numbers of low metallicity cool stars are to be expected on the blue side of the B3V line.  As well as these contaminants, eclipsing and other binaries are likely to appear in the selection region.  Of course, some binaries will include OB components.  

Our starting point is the Q3-bright catalogue, subject to the additional requirements that $g$ and $r_b$ errors are $<0.01$ mag., and that $(g-r_b)^c_V < 2.4$ mag. (to cut out stars whose $u$ magnitudes may be affected by red leak).
The initial selection on the blue side of the B3\,V reddening line produces 7190 OB candidates.  To be selected, the $(u-g)$ colour of an object must be displaced from the cut line by more than $5\sigma$, where $\sigma$ is the quadrature sum of the appropriately-weighted photometric errors (see Appendix~\ref{sec:blue_sel}).

\begin{figure}
\begin{center}
\includegraphics[width=0.8\columnwidth]{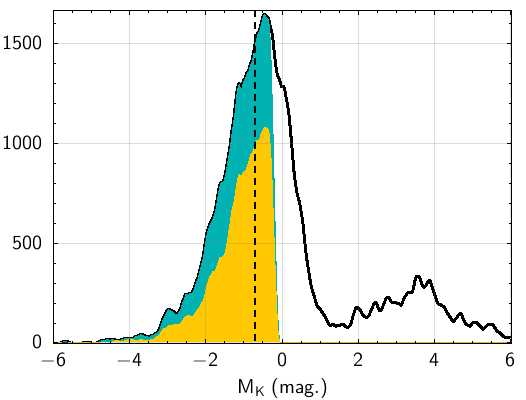}
\caption{Distribution of absolute $K$ magnitudes for stars selected above the B3\,V reddening line in the $ugr$ plane (black outline). The absolute $K$ magnitude of a B3V star, $-$0.71 mag., is marked by the dashed black line.  The cyan filled histogram shows the distribution after cutting at $M_K = -0.19$, while the yellow, superposed, is the same after rejecting sources with lower quality or negative parallaxes (see text).  KDE representation is used with Epanechnikov smoothing.
}
\label{fig:OB_MK}
\end{center}
\end{figure}

\begin{figure}
\begin{center}
\includegraphics[width=0.9\columnwidth]{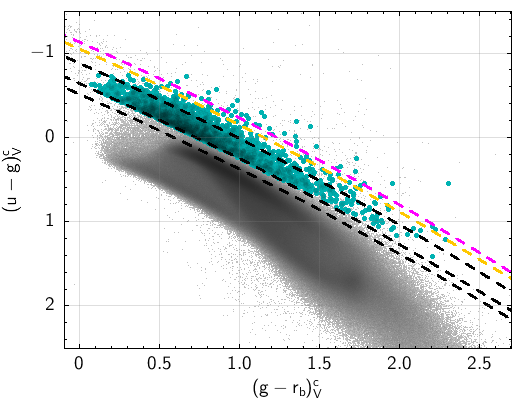}
\includegraphics[width=0.9\columnwidth]{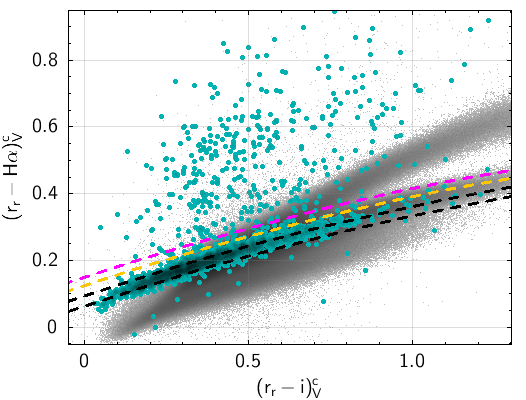}
\caption{Colour-colour diagrams for the 2568 blue-selected OB-star candidates for which the estimated $M_K$ is less than $-0.19$ (B6\,V).  Top panel: the $ugr$ plane, with the selected stars shown in cyan (as in Fig.~\ref{fig:OB_MK})  The black dashed lines are B5-7\,V, B3\,V and B1\,V reddening lines, while the yellow and magenta dashed lines are respectively the B1\,I and O9\,V reddening lines The main stellar locus is shown in grey in the background.  Bottom panel: the $riH\alpha$ plane.  Colourings and lines included are as for the $ugr$ plane, except that the B3\,V reddening line is omitted.  
}
\label{fig:OB_ccds}
\end{center}
\end{figure}

The selection can be cleaned by imposing a further criterion that separates out higher luminosity stars.  To make this separation, while minimising the extinction uncertainties, we turn to a determination of $M_K$, the $K$-band absolute magnitude.  To access the apparent $K$ magnitude needed, we cross-match the blue-selected candidates with 2MASS.  At this point we lose over 1000 candidates whose NIR magnitudes are too faint for the 2MASS catalogue (typically stars with $r \gtrsim 18$).  Few, if any, of the stars lost will be massive OB stars.  To obtain object distances, we cross-match the list with Gaia DR3, and invert parallax using the same EDSD method as used in Section~\ref{sec:astars}.  The astrometry is assured by requiring that a parallax exists and has been obtained via a 5-parameter solution, and that the Gaia DR3 parameter, RUWE, is $< 1.4$.  Similarly, to be sure of good-quality infrared photometry, we only accept A-quality 2MASS $K$ magnitudes.  After all constraints have been applied, the number of candidates remaining is 4097. 

Whilst extinction in the $K$ band is much less than in the visible, it is not zero and should be taken into account. A rough evaluation of $A_K$ can be obtained using the $(r_r - i)^c_V$ colour excess \citep[assuming a B3V intrinsic colour -- see Table A2 in][]{Drew2014}, and $A_K/A_0 \simeq 0.1$.  For contaminating cool stars, these estimates will be too high.  In a few instances, this will leave them in the 'clean' sample.  The $A_K$ estimates applied have a median of 0.22 mag., with a 95th percentile of 0.51 mag.

\begin{figure*}
\begin{center}
\includegraphics[width=1.95\columnwidth]{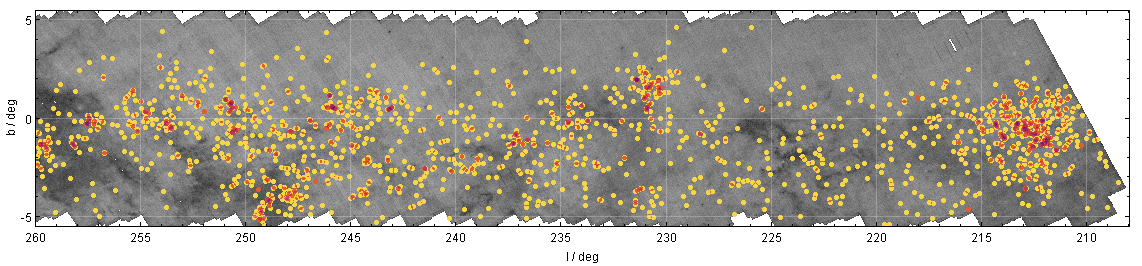}
\caption{Map of the region showing the sky positions of the OB-star candidates remaining after applying $M_K$ and parallax quality cuts (1606 stars).  The Q3-bright catalogue as modified for OB selection is included as a density modulated background in grey.  The OB candidates are coloured according to density, also, with redder colours showing where there is more crowding.  
}
\label{fig:OB_map}
\end{center}
\end{figure*}

\begin{figure}
\begin{center}
\includegraphics[width=0.78\columnwidth]{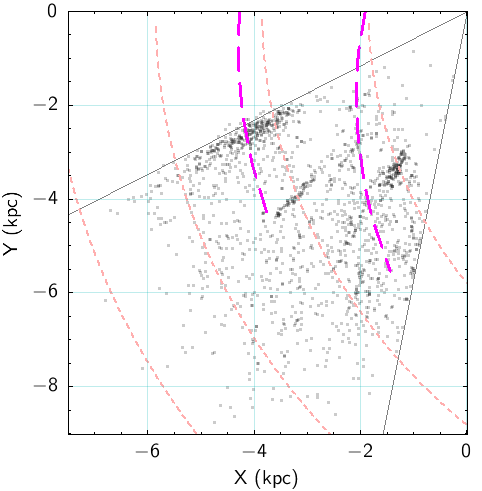}
\caption{Equatorial plane projection showing the positions of the final {\sc rplx} $> 5$ list of OB candidates (1606 stars). The OB candidates are presented in greyscale according to density.  The Sun's position is at the origin (top right corner), with the Galactic centre at (8.15,0) in the adopted Cartesian system.  The grey lines mark Galactic longitudes, $210^{\circ}$ and $260^{\circ}$. The pink dashed lines are lines of constant Galactocentric radius ($R = 10$, 12, 14 and 16 kpc).  The dashed line segments in magenta are fits to the mean loci of the Perseus and Outer Arms due to Reid et al. (2019).     
}
\label{fig:OB_plane_map}
\end{center}
\end{figure}

\begin{figure}
\begin{center}
\includegraphics[width=0.9
\columnwidth]{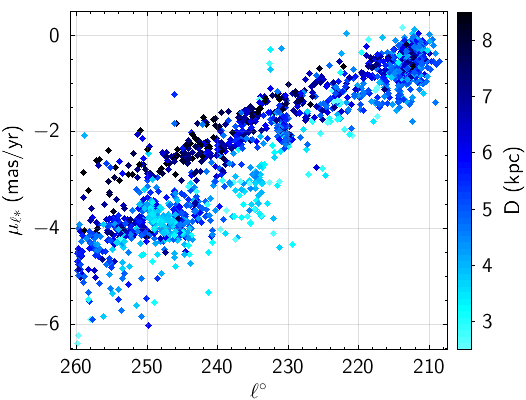}
\caption{Longitudinal proper motion component, computed from Gaia DR3 data, as a function of Galactic longitude.  The subset of OB stars with $RPLX > 5$ is shown.  The datapoints are coloured for distance as determined from parallax by the EDSD method.
}
\label{fig:OB_PMs}
\end{center}
\end{figure}

After extinction correction, the OB candidates with $M_K$ estimates are distributed as shown in Fig.~\ref{fig:OB_MK} (black outline).  The peak of the distribution falls near expectation for B3\,V stars ($M_K < -0.71$)\footnote{The absolute K magnitudes cited here are taken from E. Mamajek's table of dwarf colours and related data, linked from https://www.pas.rochester.edu/~emamajek/.}.  The 80th percentile of the distribution falls just below $M_K \simeq 0.65$ ($\sim$B9.5\,V) indicating most of the candidates are at least B stars. For now, we place our cut a little beyond the peak at $M_K = -0.19$ (B6\,V) for which we have a B5-7 reddening line to compare with our results.  There are 2568 stars brighter than the chosen $M_K$ cut.  The separation in $T_{{\rm eff}}$, as provided in Gaia DR3 for 87\% of the accepted candidates, is good in that their median is $\sim15$~kK, dropping to $\sim6.5$~kK for those excluded.  The cool-star fraction, lingering in this intrinsically bright selection, is likely to be in the region of 0.05.

It is instructive to see the translation of the OB candidates' colours from $ugr$ into the $riH\alpha$ plane.  This is shown in Fig.~\ref{fig:OB_ccds}.  In the $riH\alpha$ plane, most of the OB stars lie in a narrow strip within the main stellar locus, with a scatter of emission line stars above them.  In most cases, the latter will be classical Be stars.  Some stars just above the main strip may be rarer instances of O stars or B supergiants.
The inclusion of mid B stars in the candidates list is apparent in $riH\alpha$ as shown by the B5-7\, V reddening line running along the lower edge of the narrow strip occupied.  However, in the $ugr$ plane  the B5-7\,V reddening line is offset by $\sim0.2$ mag below the selection line.  Metallicity effects in $ugr$ are undoubtedly part of the explanation for this gap.  In the outer disk there will be B stars with reduced [Fe/H] and bluer $(u-g)$ than their solar metallicity counterparts.  Unfortunately, we do not have to hand theoretical reddening lines for low-metallicity B type stars to test this expectation.  Photometric error, particularly in $u$, may also push later-type B stars over the selection boundary.  The random errors assigned to the brighter stars ($u < 18$) involved here could be very optimistic ($<0.005$~mags -- see Fig.~\ref{fig:errors_by_filter}).  In $u$, the flat field is not as well-exposed as it is in the other bands, introducing uncertainty that can dominate the error budget, without showing up in the pipeline-reported error.

Only a few O stars are to be expected in the selection because of the limited cumulative extinction in the third quadrant.  O9.5 V stars, with $M_V \simeq -4.0$ ($M_K = -3.1$), and typical extinction of $\sim$2 visual magnitudes, just become faint enough for inclusion here if located at a distance of around 10~kpc or more. Candidates in our list with $M_K < -3.1$ associate with higher extinctions of 3 to 5 magnitudes, and distances in excess of $\sim5$~kpc.  More O stars will be detected as unsaturated sources behind the higher extinctions characteristic of the 4th and 1st Galactic quadrants.

Our blue selection for OB stars is critically dependent on the +0.26 mag. calibration offset applied to $u$ data.  If this offset is reduced, more stars are selected, and contamination rises.  However, it turns out that contamination declines slowly as a fraction of the total selection when the offset is increased.  We have measured this by looking at the variation of median Gaia DR3 $T_{{\rm eff}}$ (available for over 80\% of the selected stars) with increasing offset and find that it rises from 7.5~kK for an offset of $+0.20$, and stabilises at around $\sim$11~kK for offsets of $\gtrsim +0.25$.  We conclude that the adoption of $+0.26$ (Table~\ref{tab:offsets}) is safe.
   
The list of 2568 stars above the imposed $M_K = -0.19$ cut are, according to their parallaxes, spread in distance from 3.4~kpc up to 10.2~kpc (5th and 95th percentiles, respectively).  Almost none of these stars can be related to the Local Arm: it is too close, even, for unsaturated mid-B stars.  So far we have placed no explicit constraint on parallax precision. To get a better picture of where these stars are, at the expense of the most distant, we now require the ratio of parallax to error, {\sc rplx}, to exceed 5.  This reduces the distance range, bringing the 95th percentile down to 7.8~kpc.  The effect this has on the $M_K$ distribution of the remaining 1606 stars is shown in Fig.~\ref{fig:OB_MK} (yellow). 

The on-sky positions of the final sample of 1606 stars are shown in Fig.~\ref{fig:OB_map}.  A feature of the distribution to note is how few are placed at $b > 3^{\circ}$.  Again, there is a bias towards lower Galactic latitudes, as noted for the emission line stars (Section~\ref{sec:emstars}). Given the large inferred distances of these stars -- with a median of 5.2~kpc -- this is likely to be related to the downwards warp of the Galactic plane: \cite{Soeding2025} have reconfirmed, via gas tracers, that the down warp begins in the 3rd Galactic quadrant and extends into the 4th \citep[see their Fig.11 and also previous work by][]{Levine2006}.  The mean Galactic latitude of these stars is $-1.0^{\circ}$, with a large standard deviation of $\pm1.9^{\circ}$. Re-expressed as spatial displacement with respect to the equatorial plane, this becomes become $z = -96 \pm195$~pc.  

The detail of how the stars are distributed in distance is demonstrated in the equatorial plane projection shown in Fig.~\ref{fig:OB_plane_map}.  Massive stars are detected out to Galactocentric radii of $\sim$14 kpc, on limiting the selection to {\sc rplx} $> 5$.  The stars inward of $\sim$2.5 kpc distance of the Sun are mostly associated with higher extinctions ($A_K \gtrsim 0.3$). Only three clusterings stand out, above a general scatter. Their elongation in Fig.~\ref{fig:OB_plane_map} reflects DR3 parallax error.  For now, we just note that the two less populous clusters in the vicinity of ($230^{\circ}.7$,$+1^{\circ}.7$) and ($248^{\circ}.7$,$-4^{\circ}.0$) respectively, can be linked to known open clusters (NGC~2414 and Haffner~15).  They lie on the inner side of the Outer and Perseus Arms, according to the fits to VLBI maser data performed by \cite{Reid2019}. 

However, the most obvious concentration of stars in both Figs.~\ref{fig:OB_map} and \ref{fig:OB_plane_map}, spans a region approaching 4 degrees across, around ($212^{\circ}.0, -0^{\circ}.6$). It is not a recognised OB association \citep[see the recent compilation by][]{Wright2020}. Its distance of $\sim5$~kpc lines it up with the Outer Arm, and its diameter converts to a spatial extent of over 300~pc. It accounts for around 300 B stars in our $RPLX > 5$ selection.  If there are O stars, they have been missed as too bright for the catalogue.  We have considered whether this association -- perhaps more properly viewed as a loose aggregation -- might be an artefact of the data or its calibration, but find no evidence it is.  
Evidence in favour of its reality includes a large arc of raised obscuration around and below the B-star clustering that can be seen in Figs.~\ref{fig:density_map}, \ref{fig:Astar_maps} and \ref{fig:OB_map}.   The open cluster, Dolidze 25, the HII region Sh 2-284 \citep{Negueruela2015} and the possible supernova remnant G213.0$-$0.6, over two degrees across \citep{Green2025,Li2025} are placed within it.  Is this collection of B stars an expanded relic of an OB association?

Proper motion data can help here: Fig.~\ref{fig:OB_PMs} shows Gaia DR3 longitudinal proper motions as a function of Galactic longitude for the final OB selection with better quality astrometry ($RPLX > 5$).  Two linear sequences suggest themselves: the first, less negative, sequence may be linked to the Outer Arm running all the way across the Q3-bright catalogue region, and the second starting around $\ell = 235^{\circ}$, offset below the first, may be the Perseus Arm.  The disappearance of the latter at $\ell < 235^{\circ}$ can be attributed to the $r = 13$ mag. bright limit: lightly-reddened B stars under 3 kpc away are brighter than this. The OB-star aggregate around $\ell \sim 212^{\circ}$ appears as a dense clump in the upper righthand corner of the plot.  Within the clustering, the dispersions of the proper motion components in Galactic longitude and latitude are approximately 0.32 and 0.26 mas yr$^{-1}$ ($\sim 6$ and $\sim 5$ km s$^{-1}$ at 5 kpc) -- plausible values for what may be a loosely bound/unbound structure.

\section{Concluding discussion}
\label{sec:discussion}

This presentation of VPHAS$+$ uniformly-calibrated point-source photometry in the third Galactic quadrant began in Section~\ref{sec:data} with an outline of the magnitude trends in random error. We showed, for the case of the $r$ band, that the pipeline error is usually optimistic, particularly at the brightest magnitudes where systematic effects dominate the error budget. At the faint $r = 19$ limit of Q3-bright uncertainties will usually be no worse than 0.01--0.02 mag in $g$, $r_b$ and $r_r$, $i$ and $H\alpha$.  Errors in $u$ are generally higher, and more dispersed both for sensitivity reasons and because extinction renders $u$ magnitudes at $r = 19$, 1--2 mag. fainter, even for intrinsically blue objects.

Galactic disk stars in the third quadrant lie behind much less extinction than do inner disk stars.  Accordingly broad-band stellar photometry covering this part of the disk should be less distorted by extinction-induced effective wavelength shifts than inside the Solar Circle.  But 'less' is not the same as no effect.  This became apparent on comparing VPHAS$+$ photometry with overlapping photometry in the DECaPS-2, PS1 and Skymapper surveys.  Extinction combines with the differences in filter transmission profiles between surveys to produce a complex pattern of non-zero relative band offsets and colour trends.  

We find the colour behaviour of VPHAS$+$ most resembles DECaPS-2 (see the $gri$ diagrams in Fig.~\ref{fig:gri_diagrams}).  The reason for this seems to be that the filters used by DECaPS are consistently redder than those used in VPHAS+.  Although the photometric scales of PS1 and DECaPs-2 have been aligned away from the Galactic plane \citep{Saydjari2023}, this alignment is degraded where these surveys overlap in the plane, in the Galactic longitude range $240^{\circ} < \ell < 245^{\circ}$. This has the consequence that we do not obtain the same $g$, $r$ and $i$ zero-point offsets between these surveys and VPHAS+.  And the offsets are different again when the comparison is made with Skymapper DR4. Such is the impact of even moderate extinction. Nevertheless, we conclude, that the present version of the VPHAS+ uniform calibration, keying off comparisons in the pipeline with APASS DR9, requires some minor zero-point adjustment (see Section~\ref{sec:phot-scale} and Table~\ref{tab:offsets}).  

The uniformity of the VPHAS+ calibration is tested by tracing $g$, $r$ and $i$ magnitude differences between VPHAS+ and each of DECaPS-2, PS1 and Skymapper DR4, as a function of sky position.  This assumes, necessarily, that the comparison datasets are themselves flat.  In this, we have to rely on the majority vote in that we find that the comparisons with PS1 and Skymapper pass as flat to within $\pm$0.01 mag, with some spatial gradient in the VPHAS+ $r$ band.  But there is an appreciably higher amplitude of variation with respect to DECaPS-2 (see Fig>~\ref{fig:uniformity}).  

In order to find the proper alignment of the $u$ and $H\alpha$ magnitude scales, we compare with synthetic stellar photometric tracks in, respectively, the $ugr$ and $riH\alpha$ planes (after correcting the $g$, $r$ and $i$ zeropoints). The offsets needed are significant: 0.26 mag for $u$ and 0.076 mag for H$\alpha$.  This has its origins in the relative shortage of viable standard star information or other calibrated survey data in these bands. 

We have also compared VPHAS$+$ $g$, $r$, $i$ and $H\alpha$ magnitudes with those from its northern plane counterpart, IGAPS.  The results are entirely satisfactory, although there appears to be a small as yet unexplained difference ($\sim$0.02~mag) in the H$\alpha$ scales.  A significant point to note is that the $riH\alpha$ planes only match well if the VPHAS+ magnitudes are colour-corrected into the PS1 system -- for the reason that IGAPS magnitudes are expressed in this system.

The second half of this paper has been about the selection of specific object types, starting with A stars (Section~\ref{sec:astars}) and ending with massive OB stars (Section~\ref{sec:OBstars}). Because early-type stars are intrinsically luminous, they represent a useful population for exploring the structure of the Galactic disk.  Combining the simple two-dimensional A-star maps presented in Fig.~\ref{fig:Astar_maps} with selected longitude strips of $A_0$ versus distance trends (Fig.~\ref{fig:A0_trend}), we teased apart the extinction contributions of the Local and Perseus Arms out to a distance of $\sim$5 kpc.  With the Perseus Arm located directly behind the Local Arm, on and below the Galactic equator, separating them is not without its challenge.  The baseline contribution to extinction from the general ISM, as opposed to the scatter of embedded dark clouds, is no more than $\Delta A_0 = 1$ in either arm. 

The results from the easily-selected A stars cross-check consistently with the high-resolution Bayestar19 map \citep{Green2019}. 
As future releases of Gaia astrometry provide higher quality parallaxes, A-star selection will increase in reach and precision, increasing the power of this simple approach to extinction mapping on kiloparsec scales.  Presently, there are sufficient numbers of A star candidates, even in the low-density 3rd quadrant, to support mapping down to the scale of a square degree. 

The OB-star selection, picking up stars mainly beyond $\sim$3 kpc, found more stars at negative latitudes. The extent of this bias among early type stars is consistent with the down-warp of the Galactic plane known mainly from gas tracers.  An interesting feature of the selection, worthy of follow-up, is the discovery of a very extended concentration of B stars around Galactic coordinates ($212^{\circ}.0$ , $-0^{\circ}.6$), seemingly in the Outer Arm, $\sim$5 kpc away.  We have seen here that selecting OB stars beyond the Solar Circle is affected by declining metallicity: essentially, later-type B stars can be as blue in $(u-g)$ as solar-metallicity B3 and earlier-type stars. A route to resolving this degeneracy is to refine the photometric selection by e.g. fitting candidates' spectral energy distributions \citep[cf. the Carina Arm study by][]{MMS2017}.  

In Section~\ref{sec:emstars} on the selection of emission line stars, attention was drawn to the problem of variability and binary-star contamination.  It creates a problem for using the $girH\alpha$ plane to select emission line objects: the contemporaneous $riH\alpha$ plane remains the better choice.  Despite the design of VPHAS$+$, intended to minimise the distortion of measured stellar colours by variability, it is still an issue, mainly because short-period binaries exhibiting light variations on timescales of under an hour are about as common as the classical Be stars and YSOs that make up most of the emission line stars. In principle, the same issue arises in selecting A and OB stars from the fringes of main stellar loci. Before now, short period binaries have been found to contaminate the blue selection of OB stars \citep{MMS2017}.  However A-star selection, at least, will be less affected given that A stars are relatively abundant.

Any survey using an observing plan that limits filter changes, allowing hours, days or even weeks to elapse between field revisits to build up colours will be impacted more by variability. The collection of data in three filters in under an hour is a critical feature of the VPHAS+ survey and is the origin of the two distinct $r$ band magnitudes, $r_r$ and $r_b$. The observing strategy for Skymapper ensured contemporaneous collection of images in all its broadband filters. The Gaia mission stands out as a rare instance of a survey where effective simultaneity across the optical range was built in.  Contemporaneous colour acquisition was not a feature of either the Pan-STARRS or DECaPS data collection strategies, and it is unlikely to be preferred in upcoming Rubin Observatory operations \citep{LSST2019,Street2023}.  

In Section~\ref{sec:SIMBAD}, cross-matches with the literature were used to expose the photometric behaviours of subluminous stars, extreme red giants and young stellar objects in the VPHAS+ survey.  Usefully,
we find that the positioning of white dwarfs in the $ugr$ plane requires a large estimated $u$ zeropoint offset compatible with the +0.26 mag. estimated in Section~\ref{sec:tracks}.   Differences between the $H\alpha$ filters deployed at the INT and VST produce photometric behaviours among the extreme red giants, that are not quite the same.  However, for the carbon stars, we notice that the $gri$ plane could provide the means for efficient selection (see Fig.~\ref{fig:LPVs_plus}).  A better picture of this and of the extreme red-giants, more broadly, will come from the inner Galactic plane (and hence the full VPHAS+ catalogue) where these stars are more common. To support YSO science, VPHAS+ data can identify active accretors via both $r_r - H\alpha$ and $u - g$ excesses, help confirm broad spectral type, and -- more subtly -- help tease out potential contaminants.

Q3-bright, the catalogue down to $r = 19$, underlying the work of this paper, is being made available through the Strasbourg Astronomical Data Centre\footnote{CDS: see https://vizier.cds.unistra.fr/viz-bin/VizieR}.  Table~\ref{tab:cat} sets out the columns it contains.  The $u$, $g$, $r_b$, $r_r$, $i$ and $H\alpha$ magnitudes listed for 11.42 million objects incorporate the best-compromise corrections obtained here.  In keeping with our finding that there is no perfect consensus on the right set of photometric zeropoints, we anticipate that third-quadrant magnitudes in the upcoming full catalogue will further minor ($\sim0.01$) mag. adjustment.  

The uniform catalogue for the entire VPHAS+ footprint is in an advanced stage of preparation (Greimel et al. in prep.).  The southern Galactic plane includes the most extreme star-field environments: the most dense and brilliant in the bulge around Baade's Window, the most sparse and locally obscured in the Aquila Rift, and the most intense emission-line nebulosity in the Carina Nebula. These present larger challenges to calibration and catalogue construction than those met here.  With the release of the full footprint, we will again make comparisons in $g$, $r$ and $i$ with DECaPS-2, PS1 and Skymapper DR4.  Because of the greater impact of both extinction and crowding, the new set of comparisons will very likely be more complex, than encountered so far.

\section*{Acknowledgements}

This paper uses data products from the ESO Telescopes at the La Silla Paranal Observatory -- specifically, data from the VST Photometric H$\alpha$ Survey of the Southern Galactic Plane and Bulge (ESO programme 177.D-3023). 
Data from the European Space Agency mission Gaia (https://www.cosmos.esa.int/gaia), processed by the Gaia Data Processing and Analysis Consortium (DPAC, https://www.cosmos.esa.int/web/gaia/dpac/consortium) have also been used.  Much of the analysis presented has made use of {\sc TopCat} \citep{Taylor2005}.  JD and RG would like to thank the Centre for Astrophysics Research at the University of Hertfordshire for continued access to their supercomputing facilities. 
RR acknowledges support from Grant RYC2021-030837-I funded by MCIN/AEI/ 10.13039/501100011033 and by “European Union Next Generation EU/PRTR”, along with partial support by the AGAUR/Generalitat de Catalunya grant SGR-386/2021 and the Spanish MINECO grant, PID2023-148661NB-I00.  

Comments provided by an anonymous referee have helped improve this paper's content.

\section*{Data Availability}

The catalogue, Q3-bright, described in this paper will be directly available through CDS. The catalogue's contents are outlined in Appendix~\ref{sec:catalogue}.  


\bibliographystyle{mnras}
\bibliography{JED_papers}

\appendix

\section{Selection of A stars}
\label{sec:Astar_selection}

\begin{figure}
\begin{center}
\includegraphics[angle=0,width=0.9\columnwidth]{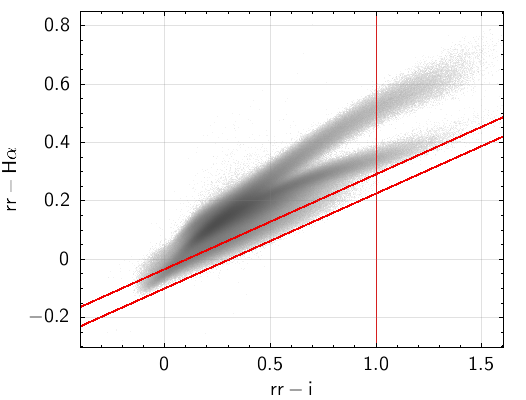}
\caption{Selection of candidate A stars from the Q3-bright via the $riH\alpha$ colour-colour plane.  The red lines mark the limits of the band from which candidate A stars are selected. The distribution of stellar colours for the Q3-bright catalogue are shown in density greyscale.     
}
\label{fig:aselection}
\end{center}
\end{figure}

Section~\ref{sec:astars} makes use of a selection of A stars.  This is how it is constructed.    
A band is placed in the $riH\alpha$ plane that identifies a zone at the foot of the main stellar locus with strongest H$\alpha$ absorption (Fig.~\ref{fig:aselection}).  The input magnitudes are in the AB system, and can be left uncorrected in the initial selection process.  Here, a pair of simple empirical linear functions define the upper and lower limits of the selection zone: their form is
\begin{equation}
r_r - H\alpha < c + 0.325(r_r-i)
\end{equation}
where $c$ is -0.034 (upper edge) or -0.1 (lower edge). Stars redder than $(r_r - i) = 1$ are excluded as these are more likely to be later-type giants.  To assure photometric quality, only stars with $r_r$ and $i$ magnitude errors less than 0.01 mag., $H\alpha$ errors less than 0.015 mag. and {\sc errbits} parameters less than 3 are considered.  This produces a list of $\sim490$k candidates.

A simple, and generous selection like this will be contaminated by cooler stars.  To estimate this and do better, the next step is to cross-match with Gaia DR3. The number of stars that fail to cross-match is negligibly small.  DR3 stellar parameters are available for 80\% of the total list (after culling $RUWE > 1.4$ objects).  Of these around a third are cooler than 7500~K.  To reduce this fraction, and to better assure parallax quality without lowering the sampling spatial density unduly, we impose a limit on acceptable parallax error, {\sc rplx} $> 2$.  This also acts to lower the cool-star contamination for the simple reason cooler stars are fainter, and astrometrically noisier, than A stars.  We estimate, from the stars with measured $T_{{\rm eff}}$, that the cut on ${\sc rplx}$ reduces the cool-star contamination to 15\%.  It leaves us with 225k stars for mapping the region - roughly one per square arcminute.


\section{Equations used in OB-star selection}
\label{sec:blue_sel}

The selection identifies stars with $(u-g)$ Vega colours that are offset to the negative side of the B3V reddening line in the $ugr$ diagram by some margin for error.   To determine if a star qualifies, it is helpful to have a polynomial in $(g-r_b)$ that tracks the B3V reddening line (that was itself determined via synthetic photometry).  The expression used here is:
\begin{equation}
(u-g)_{B3V} = -0.635 + 0.83(g-r_b) + 0.061(g-r_b)^2
\end{equation}
The excess to be computed is: 
\begin{equation}
(u-g) - (u-g)_{B3V}
\end{equation}
To qualify as a candidate OB star, this difference needs to be negative and 5 times the magnitude of the cumulative error ($\sigma$), contributed by the $u$, $g$ and $r_b$ magnitudes, in this difference.  If the gradient of the reddening line at the point of calculation of $\sigma$ is $m$, then:
\begin{equation}
\sigma = [\epsilon_u^2 + (1 + m)^2\epsilon_g^2 + m^2\epsilon_{r_b}^2]^{1/2}
\end{equation}
where $\epsilon$ represents magnitude error. 
The quadratic form of the fit to the B3V reddening line makes it easy to determine the numerical value of $m$ as a function of $(g-r_b)$, at any point along its length. By differentiation, it will be:
\begin{equation}
m = 0.83 + 0.122(g-r_b)
\end{equation}
In summary, the selection criterion is:
\begin{equation}
[(u-g) - (u-g)_{B3V}]/\sigma < -5 \ \ .
\end{equation}

\section{Content of the Q3-bright catalogue} 
\label{sec:catalogue}

The columnns making up the Q3-bright catalogue are identified in Table~\ref{tab:cat}.  The catalogue provides magnitudes in both the AB and Vega systems.  The last three columns are taken from the SIMBAD database for the 88k sources that have SIMBAD cross-matches  The catalogue may be accessed via the CDS xservice, Vizier.

\begin{table}
\caption{Q3-bright catalogue columns.}
\label{tab:cat}
\begin{tabular}{llcl}
No & Column name & Units & Description \\ 
\hline
1 & RA\_2000 & degrees &  Right Ascension J2000 \\ 
2 & DEC\_2000 & degrees &  Declination J2000 \\ 
3 & $\ell$ & degrees &  IAU 1958 Galactic longitude \\
4 & $b$  & degrees &  IAU 1958 Galactic latitude \\
5 & mergedClass &    &  Band-merged morphological class \\
6 & i\_AB & mag &  $i$ band AB magnitude \\
7 & i\_V & mag & $i$ band Vega magnitude  \\
8 & err\_i & mag &  $i$ band photometric uncertainty \\ 
9 & Class\_i &    & $i$ morphological class \\
10 & Av\_conf\_i &     & $i$ band confidence level \\
11 & errbits\_i &     & $i$ band cumulative warning flag   \\
12 & Ha\_AB & mag &  H$\alpha$ band AB magnitude \\
13 & Ha\_V & mag & H$\alpha$ band Vega magnitude \\
14 & err\_Ha & mag &  H$\alpha$ band photometric uncertainty \\ 
15 & Class\_Ha &    & H$\alpha$ morphological class \\
16 & Av\_conf\_Ha &     & H$\alpha$ band confidence level \\
17 & errbits\_Ha &     & H$\alpha$ band cumulative warning flag  \\
18 & rr\_AB & mag &  red-OB $r$ band AB magnitude \\
19 & rr\_V & mag & red-OB $r$ band Vega magnitude \\
20 & err\_rr & mag &  red-OB $r$ band photometric uncertainty \\ 
21 & Class\_rr &    & red-OB $r$ morphological class \\
22 & Av\_conf\_rr &     & red-OB $r$ band confidence level \\
23 & errbits\_rr &     & red-OB $r$ band cumulative warning flag \\
24 & rb\_AB & mag &  blue-OB $r$ band AB magnitude \\
25 & rb\_V & mag & blue-OB $r$ band Vega magnitude  \\
26 & err\_rb & mag &  blue-OB $r$ band photometric uncertainty \\ 
27 & Class\_rb &    & blue-OB $r$ morphological class \\
28 & Av\_conf\_rb &     & blue-OB $r$ band confidence level \\
29 & errbits\_rb &     & blue-OB $r$ band cumulative warning flag \\
30 & g\_AB & mag &  $g$ band AB magnitude \\
31 & g\_V & mag & $g$ band Vega magnitude \\
32 & err\_g & mag &  $g$ band photometric uncertainty \\ 
33 & Class\_g &    & $g$ morphological class \\
34 & Av\_conf\_g &     & $g$ band confidence level \\
35 & errbits\_g &     & $g$ band cumulative warning flag   \\
36 & u\_AB & mag &  $u$ band AB magnitude \\
37 & u\_V & mag & $u$ band Vega magnitude \\
38 & err\_u & mag &  $u$ band photometric uncertainty \\ 
39 & Class\_u &    & $u$ morphological class \\
40 & Av\_conf\_u &     & $u$ band confidence level \\
41 & errbits\_u &     & $u$ band cumulative warning flag   \\
42 & main\_id &    & SIMBAD main identifier \\
43 & main\_type &    & SIMBAD main object type \\
44 & sp\_type &    &  SIMBAD spectral classification \\
\hline
\end{tabular}
\end{table}

\end{document}